\documentclass[acmsmall,screen,nonacm]{acmart}
\usepackage{listings}
\usepackage{tikz}
\usetikzlibrary{positioning,arrows.meta,fit,calc}
\ifdefined\lBrack
  \providecommand{\llbracket}{\lBrack}
  \providecommand{\rrbracket}{\rBrack}
\fi
\setcopyright{none}
\newif\ifcodehighlight
\codehighlightfalse
\ifcodehighlight
\fi
\newcommand{\fwd}[1]{\langle #1]}
\newcommand{\bwd}[1]{[#1\rangle}
\newcommand{\hole}{\square}
\newcommand{\D}{\mathcal D}
\newcommand{\code}[1]{\textnormal{\texttt{#1}}}
\newcommand{\kamot}{\textsc{Kamot}}
\newcommand{\nd}[1]{\ensuremath{\mathsf{#1}}}
\newcommand{\seek}{\mathsf{seek}}
\newcommand{\rise}{\mathsf{rise}}
\newcommand{\emit}{\mathsf{emit}}
\newcommand{\scan}{\mathsf{scan}}
\definecolor{trailink}{HTML}{4B5D73}
\definecolor{trailfill}{HTML}{E8EDF2}
\definecolor{childink}{HTML}{5B7468}
\definecolor{childfill}{HTML}{EEF2EF}

\begin{document}
\title{Datatypes In, Zippers Out!}
\author{Smayan Agarwal}
\email{smayan@fplaunchpad.org}
\affiliation{\institution{FP Launchpad, IIT Madras}
  \city{Chennai}
  \country{India}}
\author{Aalok Thakkar}
\email{thakkar@ashoka.edu.in}
\affiliation{\institution{Vachani School of Advanced Computing, Ashoka University}
  \city{Sonipat}
  \country{India}}
\renewcommand{\shortauthors}{Smayan Agarwal and Aalok Thakkar}

\begin{abstract}
A zipper represents a focused position in an algebraic data structure together
with the context needed to reconstruct the whole. Datatype derivatives
determine one-hole context types, but do not by themselves choose a zipper's
concrete representation or specify navigation costs and demands. We present
\kamot{}, a tool that converts regular algebraic datatype
declarations in a restricted Haskell fragment into zipper implementations
with operational contracts. Navigation, replacement, and reconstruction are
proved correct against a reference traversal. The generated code uses typed derivative contexts
and a finite stack of frames (a trail), with the nearest frame on top.
It supports mutual recursion and regular containers inside containers.

A compositional analysis derives sound bounds on transitions
and constructor demands for certified traversal directions. These bounds
are affine in trail depth (the number of saved frames, which also counts
earlier siblings) and guarantee
termination of each move on productive, potentially infinite (coinductive)
inputs. For inhabited families the move analysis is complete for the
generated traversal: when no move certificate is emitted, no bound depending
only on trail depth exists for its moves, and some productive input makes a
move diverge. Accounting for each frame's
return and yield costs gives sharper bounds; for rose trees, the constructor-demand
bound depends only on source depth. We prove that the generated clauses
preserve results and the reference order of constructor matches, including
finite prefixes of divergent searches. Lean~4 verifies simulation, bounds,
and completeness for finite values of single self-recursive datatypes, in both
directions; for these, the prefix analysis reduces to a syntactic guard. It
also constructs a productive infinite input with a reachable divergent forward
move.

On 19 datatype models and 81 generated families, every generated module
passes Haskell type checking, and independent checkers validate certificates,
traces, and bounds. In a prefetching workload, starting only searches whose
bounds fit a budget answers every local query, as dynamically fuelled search
does, and avoids discarded source evaluations on long returns, but its
conservative bounds can lose answers that fuelled search delivers.
\end{abstract}
\maketitle

\section{Introduction}
\label{sec:question}
A zipper represents a position in an algebraic data structure together with
its surrounding context~\cite{Huet1997}. It lets a program inspect a value,
replace it, and continue navigating without searching again from the root.
The surrounding structure is retained, so the program can reconstruct the
updated whole while older versions remain available. These operations are
useful when an editor moves through a syntax tree, or a consumer pauses and
resumes a traversal of lazily produced data.

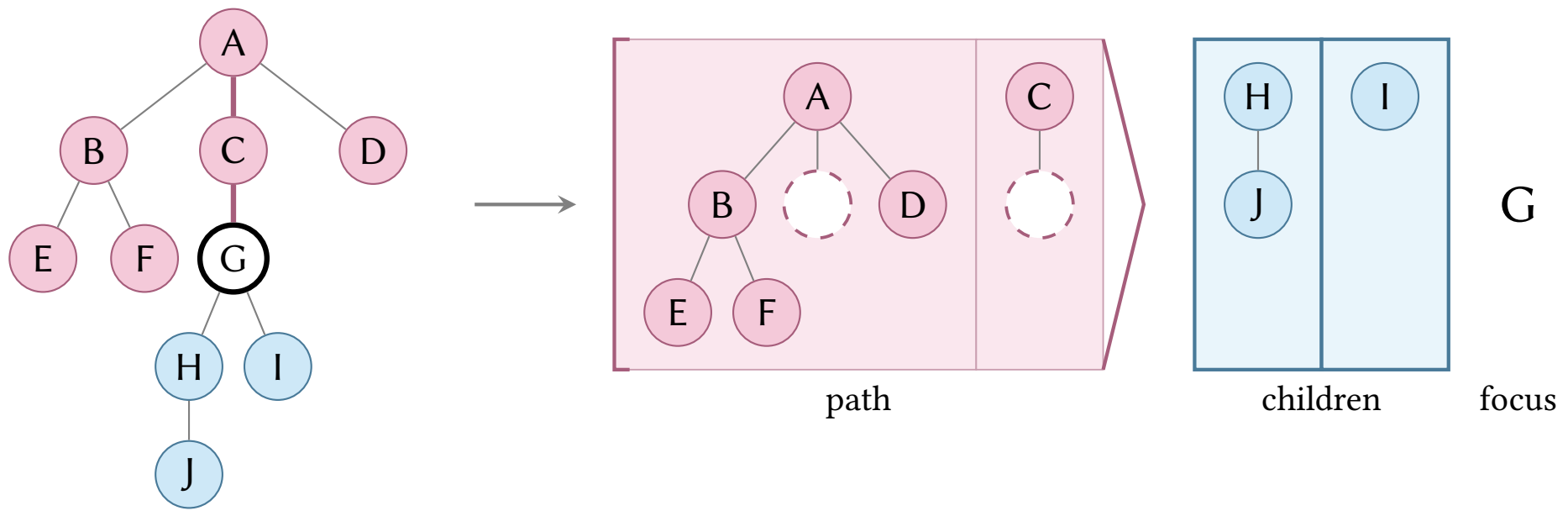
\begin{figure}[t]
\centering
\definecolor{pathdraw}{HTML}{A65D7B}
\definecolor{pathfill}{HTML}{F4C9D9}
\definecolor{kiddraw}{HTML}{497A99}
\definecolor{kidfill}{HTML}{CEE8F7}
\begin{tikzpicture}[font=\small,
  treenode/.style={circle,draw=gray,inner sep=0pt,minimum size=15pt},
  pathnode/.style={treenode,draw=pathdraw,fill=pathfill},
  kidnode/.style={treenode,draw=kiddraw,fill=kidfill},
  focusnode/.style={treenode,very thick,draw=black,fill=white},
  holenode/.style={treenode,dashed,draw=pathdraw,line width=.7pt,fill=white},
  valuenode/.style={inner sep=0pt,font=\Large},
  treeedge/.style={draw=gray},
  pathedge/.style={draw=pathdraw,very thick},
  pathcell/.style={draw=pathdraw!60,fill=pathfill!45,line width=.35pt},
  kidcell/.style={draw=kiddraw,fill=kidfill!45,thick},
  pathcap/.style={thick,pathdraw},
  lbl/.style={font=\footnotesize,below}]
\node[pathnode]  (A) at (0,0)        {$\nd{A}$};
\node[pathnode]  (B) at (-1.1,-.85)  {$\nd{B}$};
\node[pathnode]  (C) at (0,-.85)     {$\nd{C}$};
\node[pathnode]  (D) at (1.1,-.85)   {$\nd{D}$};
\node[pathnode]  (E) at (-1.5,-1.7)  {$\nd{E}$};
\node[pathnode]  (F) at (-.7,-1.7)   {$\nd{F}$};
\node[focusnode] (G) at (0,-1.7)     {$\nd{G}$};
\node[kidnode]   (H) at (-.35,-2.55) {$\nd{H}$};
\node[kidnode]   (I) at (.35,-2.55)  {$\nd{I}$};
\node[kidnode]   (J) at (-.35,-3.4)  {$\nd{J}$};
\foreach \p/\q in {A/B,A/D,B/E,B/F,G/H,G/I,H/J}{\draw[treeedge] (\p)--(\q);}
\draw[pathedge] (A)--(C); \draw[pathedge] (C)--(G);
\draw[->,>=stealth,thick,gray] (1.9,-1.275)--(2.7,-1.275);
\begin{scope}[shift={(0,-.425)}]
\draw[pathcell] (3.0,-2.15) rectangle (5.85,.45);
\draw[pathcell] (5.85,-2.15) rectangle (6.85,.45);
\draw[pathcap] (3.12,-2.15)--(3.0,-2.15)--(3.0,.45)--(3.12,.45);
\draw[pathcap] (6.85,-2.15)--(7.17,-.85)--(6.85,.45);
\begin{scope}[shift={(4.6,0)}]
  \node[pathnode] (fA) at (0,0)        {$\nd{A}$};
  \node[pathnode] (fB) at (-.75,-.85)  {$\nd{B}$};
  \node[holenode] (h1) at (0,-.85)     {};
  \node[pathnode] (fD) at (.75,-.85)   {$\nd{D}$};
  \node[pathnode] (fE) at (-1.1,-1.7)  {$\nd{E}$};
  \node[pathnode] (fF) at (-.4,-1.7)   {$\nd{F}$};
  \foreach \p/\q in {fA/fB,fA/h1,fA/fD,fB/fE,fB/fF}{\draw[treeedge] (\p)--(\q);}
\end{scope}
\node[pathnode] (fC) at (6.35,0)    {$\nd{C}$};
\node[holenode] (h2) at (6.35,-.85) {};
\draw[treeedge] (fC)--(h2);
\node[lbl] at (4.925,-2.15) {path};
\draw[kidcell] (7.57,-2.15) rectangle (8.57,.45);
\draw[kidcell] (8.57,-2.15) rectangle (9.57,.45);
\node[kidnode] (cH) at (8.07,0)    {$\nd{H}$};
\node[kidnode] (cJ) at (8.07,-.85) {$\nd{J}$};
\node[kidnode] (cI) at (9.07,0)    {$\nd{I}$};
\draw[treeedge] (cH)--(cJ);
\node[lbl] at (8.57,-2.15) {children};
\node[valuenode] at (10.12,-.85) {$\nd{G}$};
\node[lbl] at (10.12,-2.15) {focus};
\end{scope}
\end{tikzpicture}
\caption{A rose tree and its zipper focused on the label $\nd{G}$: path, children,
focus. The path holds one crumb per ancestor, drawn root first; a crumb keeps
the ancestor's label and its other children around a hole (dashed). The
path's pointed end, accessed first, exposes $\nd{C}$'s crumb. The bounds in
the text count frames. The path has five: $\nd{A}$'s node and the list cells
holding $\nd{B}$ and $\nd{C}$, then $\nd{C}$'s node and the cell holding
$\nd{G}$.}
\label{fig:rose-zipper}
\Description{On the left, a tree with root A, children B, C and D, where B
has children E and F, C has child G, G has children H and I, and H has child
J. G is outlined. On the right, its zipper: a path of two crumbs, A with B, a
hole and D, then C with a hole; the children H with J, and I; and the label
G.}
\end{figure}

Consider a rose tree, in which each node holds a label and a list of
children:
\begin{lstlisting}
data Tree a = Node a [Tree a]
\end{lstlisting}
Figure~\ref{fig:rose-zipper} shows a zipper focused on the label $\nd{G}$ of
a ten-node tree (node labels are set in sans serif). It keeps three parts. The \emph{path} has one \emph{crumb} per
ancestor: the ancestor's label and the siblings on either side of the child
that was entered, with a hole in its place. The crumb nearest the focus,
$\nd{C}$'s, sits at the accessible end of the path. Returning to $\nd{C}$
uses that crumb directly, filling its hole with the rebuilt subtree at $\nd{G}$.
Within a crumb, the nearest earlier sibling and the nearest later sibling
are likewise the ones exposed. The \emph{children} component is $\nd{G}$'s
child list, and the \emph{focus} is its label. Together with the path they
reconstruct the tree.

Datatype derivatives compute one-hole context types~\cite{McBride2001,Abbott2005};
the crumb above is a value of one. McBride's
dissection goes further, deriving executable navigation and tail-recursive
folds by distinguishing processed results from pending
inputs~\cite{McBride2008}. We build on these constructions to derive
\emph{operational contracts}: bounds on the work done and the source
constructors forced between one focused label (an \emph{observation}) and the next.

Such guarantees matter even though the zipper reaches the nearest crumb in
one step. In Figure~\ref{fig:rose-zipper}, a forward move from
$\nd{G}$ inspects the first cell of its child list and the node it holds,
then focuses $\nd{H}$, leaving the other children unforced. A forward move from $\nd{I}$ must instead
return through $\nd{G}$ and $\nd{C}$ before it reaches $\nd{D}$, and in
general may cross several empty child lists. On an infinite input, one search can
fail to return even though every constructor it requests is produced in
finite time. A client deciding whether to start a search needs a bound on
that search, which the type of the saved state does not provide.

Whether a move is cheap depends on the datatype. In the rose tree every
node starts with its label, so when \code{next} enters a nonempty child
list, the first node it meets yields a label at once. Seeing this requires
looking through both constructors, since a list cell has no label of its
own. Now add a constructor \code{Silent [Tree a]}: a node with children but
no label. A search that enters a chain of silent nodes keeps descending
without finding a label, and on an infinite chain it never returns. The same construction gives both datatypes zippers, but only the first guarantees a bounded \code{next}.

Our approach derives the traversal and its analysis from the same ordered
datatype description. Differentiation produces local contexts for recursive
children, stored as records called \emph{frames}; a finite stack of frames,
the \emph{trail}, keeps the nearest one accessible.
A move is a search that runs generated clauses, one per \emph{transition}
(entering a constructor, resuming after a field, or returning through a frame).
It either \emph{yields} an element or \emph{returns} without one. The analysis composes their work and demand summaries
across mutually recursive types and container fields, then along the saved
trail. A search that must yield before returning to a saved frame can be
bounded from its current position alone; one that may return also depends on
the saved frames. The analysis guarantees the following.
\begin{quote}
For every direction (forward or backward) with an emitted \emph{move certificate},
a move from a cursor with $h$ saved frames
reaches its next element or the boundary within work and constructor-demand
bounds affine in $h$, provided each demanded constructor is produced in
finite time (Theorem~\ref{thm:family-prefix}).
\end{quote}
For the rose tree, the bounds are $4+3h$ transitions and $2+2h$ constructor
demands; $h$ counts frames, for list cells as well as tree nodes, so
it grows with the number of earlier siblings on the path as well as with
depth; $h=5$ at $\nd{G}$ in Figure~\ref{fig:rose-zipper}. With
\code{Silent}, the analysis emits no move certificate, and no bound depending
only on $h$ exists (Theorem~\ref{thm:completeness}). The guarantee applies to
each observation, without requiring the input to be finite or its size to be
known. It does not make depth-first traversal fair: an infinite subtree can
still keep its later siblings from ever being visited.

A client can compare a move's bounds with its work and demand budgets before
forcing any source constructor: it can start the move when the bounds fit, or
defer it when an edit may soon cancel the query. Resumable execution splits a
longer or uncertified search into slices of bounded length. The bounds count
structural steps; evaluating one source thunk may still take arbitrary time,
and a conservative bound can defer a search that would have been cheap.

We make three contributions:
\begin{itemize}
\item \emph{Compositional contracts.} Per-type summaries of search work and demand compose
across mutually recursive types and regular containers to certify the work,
source demand, and conditional termination of each move
(Theorem~\ref{thm:family-prefix}). Charging each saved frame its own outcome
sharpens the bound (Theorem~\ref{thm:trail-sum}). For inhabited families
(every sort occurs in a finite value and has a finite value containing an
element) the move analysis is complete for the generated traversal: it emits
no move certificate exactly when no bound
depending only on $h$ exists (Theorem~\ref{thm:completeness}).
\item \emph{A compilation account.} Ordered derivative records and indexed
trails represent finite paths, including paths within infinite values.
Generated operations preserve the reference traversal's result and ordered
demand trace, including every finite prefix of a divergent search
(Theorems~\ref{thm:family-trail}, \ref{thm:lowering}, \ref{thm:entry-contract},
and~\ref{thm:move-contract}). The same construction supports a resumable
backend for eager folds (Section~\ref{sec:folds}).
\item \emph{An implementation and evaluation.} We implement this approach
in \kamot{}, which emits specialised Haskell and machine-readable contracts.
Independent tools check the certificates and execution traces. We evaluate
certificate precision, admission decisions, compiler scaling, and runtime
costs against handwritten and generic implementations
(Section~\ref{sec:applications}). In the measured family workloads, generated cursors match
the handwritten indexed stack's allocation, with overlapping timing ranges.
\end{itemize}

\section{The generated cursor interface}
\label{sec:overview}
\kamot{} compiles the rose-tree declaration of Section~\ref{sec:question}
into a Haskell module. The cursor constructors stay private. Specialised to
a rose tree, its main operations are
\begin{lstlisting}
openFirst, openLast :: Tree a -> Maybe (Zipper a Tree)
next, previous      :: Zipper a Tree -> Maybe (Zipper a Tree)
focus               :: Zipper a Tree -> a
replace             :: a -> Zipper a Tree -> Zipper a Tree
closeTree           :: Zipper a Tree -> Tree a
\end{lstlisting}
\code{openFirst} and \code{openLast} focus the first and last label in
traversal order: depth first, following constructor field order, so each
label precedes its children. \code{next} and \code{previous} move one label
along that order and return \code{Nothing} past either end. \code{focus} and
\code{replace} read and overwrite the focused label and touch only the
cursor record. \code{closeTree} rebuilds the whole tree. A client can
therefore stream the labels, pausing, resuming, or editing between calls,
without restarting from the root. On the tree of
Figure~\ref{fig:rose-zipper}, \code{openFirst} focuses $\nd{A}$, and repeated
\code{next} visits $\nd{B},\nd{E},\nd{F},\nd{C},\nd{G},\nd{H},\nd{J},\nd{I},\nd{D}$.

Each crumb is a value of a datatype derivative. Write $X$ for labels, $T$
for trees, and $\mathit{List}\,T$ for lists of trees. Treating the list informally as a field type (Section~\ref{sec:source}
makes this precise by lowering), one crumb of the path of a rose tree has type
\[
 X\times\mathit{List}\,T\,\hole\,\mathit{List}\,T,
\]
which keeps the fields in order: the label, the earlier siblings, the
hole~$\hole$ where the path continues, and the later siblings. The terminal
context has type $1\,\hole\,\mathit{List}\,T$, a hole for the focused label
followed by its children, and the focus adds an $X$.
Differentiation (written $\D$; Section~\ref{sec:source}) computes these
types at compile time; their values carry saved fields at run
time~\cite{McBride2008}. Generated code stores these contexts, passes them
between operations, and pattern matches on them to resume traversal.

\kamot{} lowers the child list to an auxiliary datatype. Each crumb becomes
several \emph{frames} (lowered records): one for the label, one per earlier
sibling, and one for the later siblings (Section~\ref{sec:family-frontend}). The most recently crossed frame is
kept at the accessible end, because it is the one the next return needs;
Section~\ref{sec:arden} derives this layout.

Because the interface focuses labels rather than subtrees, \code{next}
searches from the terminal context until it finds another label or reaches
the boundary. Removing the nearest frame takes constant structural work
(Section~\ref{sec:cost}), and the trail stays finite even when saved subtrees
are infinite; reaching the next label may take longer, and
Section~\ref{sec:infinite} bounds that search across the tree and list types.

\section{Compiling ordered datatype descriptions}
\label{sec:source}
\kamot{}'s intermediate representation (IR) describes a \emph{family} of $m$ mutually recursive datatypes, called \emph{sorts}.
Each sort is a finite sum of finite ordered products:
\[
 \Phi_i(X,\mathbf Y)=\sum_{c\in\mathcal C_i}\prod_{\ell=0}^{a_c-1}E_{c,\ell},
 \qquad E_{c,\ell}::=X\mid Y_j\mid V.
\]
Here $\mathcal C_i$ is the set of constructors of sort $i$ and $a_c$ is the
number of fields of constructor $c$. The element parameter is $X$;
$\mathbf Y=(Y_1,\ldots,Y_m)$ names recursive child sorts, and $V$ ranges
over field types independent of $X$ and all $Y_j$. A summand's degree in
$\mathbf Y$ is its number of recursive fields, and any degree is allowed.
Constructor names distinguish summands and field indices
distinguish factors. Products are ordered, so for bodies $F_1,F_2$ the products $F_1\times F_2$ and $F_2\times F_1$ are distinct, and the syntax is not quotiented by type isomorphism. An empty product
gives a nullary constructor and an empty sum an empty type. When $m=1$,
every recursive field refers to the datatype itself, as in a cons list; we
call such datatypes \emph{one-sort}.

The declaration frontend accepts one shared parameter \code{a} and
the field grammar
\[
 \tau::=\code{a}\mid T_i\,\code{a}\mid V\mid[\tau]\mid
          \code{Maybe}\,\tau\mid\code{Either}\,\tau\,\tau\mid U\,\tau .
\]
Each $U$ is a user-declared unary, strictly positive regular container.
Regular recursion reuses its parameter unchanged. Strict positivity here
permits recursive occurrences through the admitted sums, products, and
containers, and excludes function fields.

Lowering instantiates each container body at the container's argument. A
declaration is accepted only when the closure of these instantiations is
finite and every recursive family occurrence stays at the shared parameter.
Container references back to root family types would change their arguments
and are rejected; so are unreachable container declarations.

Lowering replaces every accepted
container application by a specialised auxiliary sort and its ordered regular
body (Figure~\ref{fig:rose-lowering}). Appendix~\ref{sec:lowering} proves that lowering preserves
traversal order and demand up to renaming. The generated module defines its
own lowered types and provides no conversion from Prelude values;
conversion work and its demands lie outside every contract.

\begin{figure}[h!]
\centering
\colorlet{srcdraw}{childink}\colorlet{srcfill}{childfill!40}
\colorlet{lowdraw}{trailink}\colorlet{lowfill}{trailfill!40}
\begin{tikzpicture}[
  codebox/.style={draw,rounded corners=2pt,line width=.5pt,
    inner xsep=6pt,inner ysep=5pt,font=\small\ttfamily,align=left},
  boxlabel/.style={font=\footnotesize,anchor=south west,inner sep=1pt}]
\node[codebox,draw=srcdraw,fill=srcfill] (src)
  {\textbf{data} Tree a = Node a [Tree a]};
\node[codebox,draw=lowdraw,fill=lowfill,right=13mm of src] (low)
  {\begin{tabular}{@{}l@{\ }l@{}}
   \textbf{data} Tree a&= Node a (List0 a)\\
   \textbf{data} List0 a&= List0Nil\\
   &| List0Cons (Tree a) (List0 a)
   \end{tabular}};
\draw[-{Stealth[length=5pt]},thick,gray] (src.east)--(low.west)
  node[midway,above,font=\footnotesize,text=black] {lowered};
\node[boxlabel,text=srcdraw] at (src.north west) {declared by the user};
\node[boxlabel,text=lowdraw] at (low.north west) {compiled by \kamot{}};
\end{tikzpicture}
\caption{Lowering the rose tree. The list field becomes the fresh sort
\code{List0}, whose constructors keep the order of \code{[]} and
\code{(:)}, with \code{List0} itself as the tail.}
\label{fig:rose-lowering}
\Description{On the left, a box containing the declaration data Tree a =
Node a [Tree a], labelled declared by the user. An arrow labelled lowered
points to a box on the right, labelled compiled by Kamot, containing data
Tree a = Node a (List0 a) and data List0 a = List0Nil | List0Cons (Tree a)
(List0 a).}
\end{figure}

We write $\mu$ for the least fixed point, allowing only finite unfoldings of
the bound recursion, and $\nu$ for the greatest fixed point, which also admits
infinite unfoldings. The rules below are for one sort, $T=\mu Y.\,\Phi(X,Y)$ with constructor set $\mathcal C$;
Section~\ref{sec:family-frontend} supplies their family indices and Section~\ref{sec:infinite} their reading at $\nu$.

We write $\D$ for the standard derivative~\cite{McBride2001,Abbott2005}.
Its rules on a constructor body are:
\[
\begin{gathered}
 \D_XX=1,\qquad \D_XY=0,\qquad \D_XV=0,\\
 \D_X(F_1+F_2)=\D_XF_1+\D_XF_2,\\
 \D_X(F_1\times F_2)=\D_XF_1\times F_2+F_1\times\D_XF_2.
\end{gathered}
\]
Here $0$ is the empty type and $1$ the unit type. The rules for $\D_Y$
swap the roles of $X$ and $Y$. Differentiating in each variable gives
\[
 K=(\D_Y\Phi)[T/Y],\qquad B=(\D_X\Phi)[T/Y].
\]
A $K$ stores the context around one recursive child; a $B$ stores the
context around one direct element. As in dissection~\cite{McBride2008}, the
product rule keeps every factor on its own side of the hole, so each summand
splits into the fields before the hole and those after it:
\[
\begin{aligned}
 K&=\sum_{c\in\mathcal C}\ \sum_{\ell:E_{c,\ell}=Y}
       L_{c,\ell}\hole R_{c,\ell},\qquad
 B=\sum_{c\in\mathcal C}\ \sum_{\ell:E_{c,\ell}=X}
       L_{c,\ell}\hole R_{c,\ell},\\
 L_{c,\ell}&=\prod_{\ell'<\ell}E_{c,\ell'}[T/Y],\qquad
 R_{c,\ell}=\prod_{\ell'>\ell}E_{c,\ell'}[T/Y].
\end{aligned}
\]
The square marks the hole; it is not a stored field. The compiler emits one
record constructor per summand, tagged by $(c,\ell)$. Its fields are exactly
the fields of $c$ other than $\ell$, in source order: the $L$ fields, then the
$R$ fields. The record's generated name encodes the tag, so it identifies the
constructor and where the hole falls between the $L$ and $R$ fields.

\section{Representing traversal histories}
\label{sec:arden}
\kamot{} stores the path of frames with the most recently crossed frame at
its accessible end (Section~\ref{sec:overview}). Write $\fwd{A}$ and
$\bwd{A}$ for forward and backward lists of $A$, with the angle marking the
accessible end:
\begin{lstlisting}
data Forward a  = FNil | FCons a (Forward a)
data Backward a = BNil | Snoc (Backward a) a
\end{lstlisting}
Both keep elements in the displayed order: $\fwd{1,2,3}$ is
\code{FCons 1 (FCons 2 (FCons 3 FNil))} and exposes $1$, while $\bwd{1,2,3}$ is
\code{Snoc (Snoc (Snoc BNil 1) 2) 3} and exposes $3$.

An element context either stops at a terminal context in $B$ or passes
through a frame $L\hole R$ of $K$ to the context of a child. Its finite-path
presentation is the path equation
$\D_XT=\mu Q.\,(B+L\times Q\times R)$~\cite{Abbott2005}, where
$L\times Q\times R$ abbreviates the sum over the summands of $K$. Each step
contains one $Q$, whatever the degree of $\Phi$ in $Y$: exactly one child
continues towards the selected occurrence, and the frame stores all other
fields on their own side of the hole.

A path lists its steps from the root down; step $k$ keeps the fields $l_k$
left of its hole and $r_k$ right of it. Rebuilding the value wraps each step
around the rest of the path, so the surrounding fields, read left to right, are
$l_1,\ldots,l_n$, then the terminal $b$, then $r_n,\ldots,r_1$. In
Figure~\ref{fig:rose-zipper}, the path to $\nd{G}$ gives $\nd{A}$ and the subtree at
$\nd{B}$ from the root step, then $\nd{C}$, then $\nd{G}$'s children, then $\nd{D}$ from the root
step again. A \emph{history} stores the same steps in a backward list, last
step exposed. For a history $H$, write $\mathit{before}(H)$ for its left
fields, first step to last, and $\mathit{after}(H)$ for its right fields,
last step to first.

\begin{lemma}[Path--trail correspondence]
\label{lem:path-factorisation}
For $L$, $R$, and $B$ independent of $Q$, there is a bijection
\[
 \mu Q.\,(B+L\times Q\times R)\cong\bwd{L\times R}\times B.
\]
It is \emph{order-preserving}: if a path corresponds to $(H,b)$, then
$\mathit{before}(H)$, $b$, $\mathit{after}(H)$ lists the context's stored fields in
their original left-to-right order. Converting a path of $n$
steps, in either direction, takes $O(n+1)$ structural work (Section~\ref{sec:cost}) and allocates
$O(n+1)$ spine cells.
\end{lemma}
\begin{proof}
See Appendix~\ref{app:proof-lem-path-factorisation}.
\end{proof}
Applied to the path equation, the lemma gives the element contexts
$\D_XT$ and the zippers $Z_T$ that focus an element:
\begin{equation}\label{eq:recursive}
 \boxed{\D_XT=\bwd{K}\times B,\qquad
        Z_T=\bwd{K}\times B\times X.}
\end{equation}
For a list, \code{data List a = Nil | Cons a (List a)}, a step has type
$X$, and the terminal context stores the rest of the list. So
\[
 Z_T=\bwd{X}\times T\times X:
\]
the history holds the elements before the focus, nearest exposed.

\subsection{Trails for mutually recursive families}
\label{sec:family-frontend}
The rose tree of Figure~\ref{fig:rose-zipper} is not one-sort: its
recursion passes through a list. Lowering (Section~\ref{sec:source}) turns
the list field into a fresh sort \code{List0}
(Figure~\ref{fig:rose-lowering}), giving the mutually recursive sorts
\code{Tree} and \code{List0}. Indexing sorts by name, their IR bodies are
$\Phi_{\code{Tree}}=X\times Y_{\code{List0}}$ and
$\Phi_{\code{List0}}=1+Y_{\code{Tree}}\times Y_{\code{List0}}$. A
path from a tree to its first child crosses
$\code{Tree}\to\code{List0}\to\code{Tree}$. Each frame records the parent
sort it reconstructs and the child sort it accepts, so a trail has a fixed
root sort and a changing current sort.

Formally, a family's finite values form its simultaneous least fixed point
\[
  T_i=\Phi_i(X,T_1,\ldots,T_m) \qquad (1\leq i\leq m).
\]
Here $i$ names a parent sort and $j$ one of its recursive child sorts.
Differentiating with respect to $Y_j$, the type of an immediate child of
sort $j$, gives $K_{ij}$: the information needed to reconstruct an $i$
after returning from a $j$. Differentiating with respect to $X$ gives
$B_i$, the terminal context for a selected element field of sort $i$:
\[
  K_{ij}=(\D_{Y_j}\Phi_i)[T_1/Y_1,\ldots,T_m/Y_m],
  \qquad
  B_i=(\D_X\Phi_i)[T_1/Y_1,\ldots,T_m/Y_m].
\]
Each summand of $K_{ij}$ again splits at its hole as $L_{c,\ell}\hole R_{c,\ell}$, for $c\in\mathcal C_i$ and $E_{c,\ell}=Y_j$,
and its record's name now also gives the sort~$i$.
For the rose tree, $K_{\code{Tree},\code{Tree}}$ and $B_{\code{List0}}$ are $0$: a node has
no tree field, and a \code{List0} cell has no label.
Finite paths form the least solution of this equation:
\begin{equation}\label{eq:family-path}
  C_i=B_i+\sum_{j}\ \sum_{c\in\mathcal C_i}\ \sum_{\ell:E_{c,\ell}=Y_j} L_{c,\ell}\times C_j\times R_{c,\ell}.
\end{equation}
For the rose tree it reads
\[
  C_{\code{Tree}}=B_{\code{Tree}}+X\times C_{\code{List0}},\qquad
  C_{\code{List0}}=C_{\code{Tree}}\times T_{\code{List0}}
    +T_{\code{Tree}}\times C_{\code{List0}}.
\]
A path from a root $r$ is thus a finite, well-sorted sequence of frames
ending at some current sort $i$, followed by a $B_i$.
Theorem~\ref{thm:family-trail} represents it by a backward trail from $r$
to $i$, nearest frame exposed:
\begin{theorem}[Indexed path--trail correspondence]
\label{thm:family-trail}
For every root sort $r$, there is a bijection, order-preserving in the sense
of Lemma~\ref{lem:path-factorisation},
\[
  C_r \cong \sum_i \mathsf{Trail}(r,i)\times B_i.
\]
Here $\mathsf{Trail}(r,i)$ is the type of finite, well-sorted frame paths
from $r$ to $i$.  Extending such a path by one $K_{ij}$ frame and removing its
nearest frame take constant structural work.  Adding the selected $X$ field
to its terminal context yields exactly one marked element occurrence in a
value of root sort $r$.
\end{theorem}
\begin{proof}
Use an accumulator whose current-sort index is part of its type.  It starts
empty at $r$.  On a summand $L_{c,\ell}\times C_j\times R_{c,\ell}$ with $c\in\mathcal C_i$, append the $i$-to-$j$
frame, which keeps both sides and its tag, and continue with current sort
$j$; on $B_i$, return the accumulator and that terminal. Frame order is
unchanged. Conversely, start with $B_i$, remove each exposed final frame,
and wrap the residual context between its $L$ and $R$ fields until the
trail is empty. These two
procedures are mutually inverse.  Each transfer touches only one outer trail and frame
constructor, while the terminal tag identifies the selected field.
\end{proof}

The generated target represents $\mathsf{Trail}(r,i)$ with a generalised
algebraic datatype (GADT) indexed by the root and current sorts. Each frame
record is indexed by the sort it rebuilds and the sort of its hole.
For example, \code{Frame\_Tree\_Node\_1} represents the child-list hole in
\code{Node} and has type \code{Frame a Tree List0}; each terminal record (\code{Body}) is indexed by its sort. For
the rose tree, \kamot{} emits
\begin{lstlisting}
data Trail a root current where
  BNil :: Trail a t t
  Snoc :: Trail a root parent -> Frame a parent child -> Trail a root child

data Zipper a root where
  Cursor :: Trail a root current -> Body a current -> a -> Zipper a root
\end{lstlisting}
\code{Snoc} extends a trail ending at a parent with a frame from that parent
to a child. A trail cannot be extended with a frame for the wrong parent,
and navigation cannot return through a mismatched frame. No runtime sort
check or cast is needed.

Before lowering, Equation~\eqref{eq:recursive} with the types of
Section~\ref{sec:overview} gives
\[
 Z_T=\bwd{X\times\mathit{List}\,T\times\mathit{List}\,T}
     \times\mathit{List}\,T\times X.
\]
\kamot{} stores each such crumb as several indexed frames (Section~\ref{sec:overview}).

\section{Generating zipper operations}
\label{sec:transitions}
For every one-sort datatype, \kamot{} emits the zipper type $Z_T$ of
Equation~\eqref{eq:recursive} as
\[
\begin{array}{@{}l@{\ \ }l@{\qquad}l@{}}
 \code{\textbf{data} Frame a}&\code{= Frame\_}c\code{\_}\ell\;\;
   E_{c,0}\cdots E_{c,\ell-1}\;\;E_{c,\ell+1}\cdots E_{c,a_c-1}
   &(E_{c,\ell}=Y)\\
 \code{\textbf{data} Body a}&\code{= Body\_}c\code{\_}\ell\;\;
   E_{c,0}\cdots E_{c,\ell-1}\;\;E_{c,\ell+1}\cdots E_{c,a_c-1}
   &(E_{c,\ell}=X)\\
 \code{\textbf{data} Backward a}&\code{= BNil | Snoc (Backward a) a}\\
 \code{\textbf{data} Zipper a}&\code{= Cursor (Backward (Frame a)) (Body a) a}
\end{array}
\]
\code{Frame} has one constructor for each constructor $c$ and recursive
field $\ell$ of $c$, and \code{Body} one for each element field. Each is $c$
with field $\ell$ removed: it keeps the fields before $\ell$ and those after
it, in source order, reading $X$ as \code{a} and $Y$ as \code{T a}. For the
list, \code{Cons} has fields $X\,Y$, so \code{Frame} is
\code{Frame\_Cons\_1 a} and \code{Body} is \code{Body\_Cons\_0 (List a)}. The first field
of \code{Cursor}, of type \code{Backward (Frame a)}, is the trail: the saved
frames from the root, nearest last.

One constructor-local rule generates the internal functions used by the
interface of Section~\ref{sec:overview}. The proofs and cost bounds of
Sections~\ref{sec:correctness}--\ref{sec:infinite} are stated over them.
The displayed clauses omit sort indices; for a family, the $\seek$ and
$\rise$ functions and the frames carry the indices of
Section~\ref{sec:family-frontend}. Call $X$
and $Y$ fields active. Given a current position $\ell$ in a constructor $c$
and a direction $d\in\{+,-\}$ (forward or backward), the compiler finds the
next active field $\ell'$ after $\ell$ in direction $d$; at constructor entry,
$\ell$ is $-1$ forward and $a_c$ backward. Each generated clause then has one
of three outcomes: yield a direct element, descend through a recursive child
after saving a frame, or return to the parent when no active field remains.

Let $\mathbf v$ be the variables bound by the constructor pattern and let
$\widehat{\mathbf v}_{\ell'}$ omit $v_{\ell'}$; $H$ is the trail. Write $k_{c,\ell'}$ and
$b_{c,\ell'}$ for the constructors \code{Frame\_}$c$\code{\_}$\ell'$ and
\code{Body\_}$c$\code{\_}$\ell'$. The entry function $\seek_d$ finds the first element of a subtree in
direction $d$; the resumption function $\rise_d$ continues in the parent
once a child is finished. The three outcomes are emitted as
\begin{equation}\label{eq:emit}
\emit_d(c,\ell)=
\begin{cases}
 \code{Just}(\mathsf{Cursor}(H,b_{c,\ell'}(\widehat{\mathbf v}_{\ell'}),v_{\ell'}))
     & E_{c,\ell'}=X,\\
 \seek_d(\mathsf{Snoc}(H,k_{c,\ell'}(\widehat{\mathbf v}_{\ell'})),v_{\ell'})
     & E_{c,\ell'}=Y,\\
 \rise_d(H,c(\mathbf v)) & \text{no such }\ell'.
\end{cases}
\end{equation}
Constant fields ($V$) are retained but never examined by this search.

The rule also produces the bodies of the public moves. A $\seek_d$ clause
matches a source constructor, saving a frame at each step down; a $\rise_d$
clause matches the nearest frame and binds the completed child to the
missing variable $v_\ell$.

The public \code{next} and \code{previous} match the cursor's terminal
record, bind the focused element to $v_\ell$, and continue after that
position; \code{openFirst} and \code{openLast} are $\seek_d$ from the empty
history. All of these return \code{Maybe (Zipper a)}. Generated code allocates a parent constructor only when traversal returns
out of it; moving to another field of the same constructor builds the next
\code{Frame} or \code{Body} record directly from the bound variables.

For the rose tree, \kamot{} emits these records, indexed by sort:
\begin{lstlisting}
data Frame a parent child where
  Frame_Tree_Node_1 :: a -> Frame a Tree List0
  Frame_List0_List0Cons_0 :: (List0 a) -> Frame a List0 Tree
  Frame_List0_List0Cons_1 :: (Tree a) -> Frame a List0 List0

data Body a current where
  Body_Tree_Node_0 :: (List0 a) -> Body a Tree
\end{lstlisting}
Using \code{Trail} and \code{Zipper} from Section~\ref{sec:family-frontend}, the forward clauses are:
\begin{lstlisting}
seekTreeF :: Trail a root Tree -> Tree a -> Maybe (Zipper a root)
seekTreeF hs (Node v0 v1) =
  Just (Cursor hs (Body_Tree_Node_0 v1) v0)

seekList0F :: Trail a root List0 -> List0 a -> Maybe (Zipper a root)
seekList0F hs List0Nil =
  riseF hs List0Nil
seekList0F hs (List0Cons v0 v1) =
  seekTreeF (Snoc hs (Frame_List0_List0Cons_0 v1)) v0

riseF :: Trail a root current -> current a -> Maybe (Zipper a root)
riseF BNil _ = Nothing
riseF (Snoc hs frame) child = resumeF hs frame child

resumeF :: Trail a root parent -> Frame a parent child -> child a -> Maybe (Zipper a root)
resumeF hs (Frame_Tree_Node_1 v0) v1 =
  riseF hs (Node v0 v1)
resumeF hs (Frame_List0_List0Cons_0 v1) v0 =
  seekList0F (Snoc hs (Frame_List0_List0Cons_1 v0)) v1
resumeF hs (Frame_List0_List0Cons_1 v0) v1 =
  riseF hs (List0Cons v0 v1)

next :: Zipper a root -> Maybe (Zipper a root)
next (Cursor hs (Body_Tree_Node_0 v1) v0) =
  seekList0F (Snoc hs (Frame_Tree_Node_1 v0)) v1
\end{lstlisting}
Except for the empty-trail boundary and sort dispatch, these clauses
instantiate Equation~\eqref{eq:emit}. The second \code{resumeF} clause returns
from a \code{Tree} child to its \code{List0} parent, changing the current sort
accordingly. Entry is \code{openFirst = seekTreeF BNil}.

The same rule, with the order of active fields reversed, generates the
backward functions \code{seekTreeB}, \code{seekList0B}, \code{riseB}, and
\code{resumeB}, and from them \code{previous} and \code{openLast}.
Moving between two element fields of one constructor replaces the
\code{Body} record and leaves the trail unchanged; constructors that
interleave elements and children need no further case. A \emph{search clause} is one outcome of Equation~\eqref{eq:emit} or the
\code{BNil} boundary clause. A call to a sort-dispatch helper such as
\code{riseF} counts as part of the outcome that makes it;
Section~\ref{sec:cost} charges its cost to that clause's structural overhead.

The closing operation, \code{close} (here \code{closeTree}), rebuilds one parent per frame:
\begin{lstlisting}
rebuild :: Frame a parent child -> child a -> parent a
rebuild (Frame_Tree_Node_1 v0) v1 = (Node v0 v1)
rebuild (Frame_List0_List0Cons_0 v1) v0 = (List0Cons v0 v1)
rebuild (Frame_List0_List0Cons_1 v0) v1 = (List0Cons v0 v1)

finish :: Trail a root current -> current a -> root a
finish BNil t = t
finish (Snoc hs frame) child = finish hs (rebuild frame child)

closeTree :: Zipper a Tree -> Tree a
closeTree (Cursor hs body x) = finishBodyTree hs body x

finishBodyTree :: Trail a Tree current -> Body a current -> a -> Tree a
finishBodyTree hs (Body_Tree_Node_0 v1) v0 =
  finish hs (Node v0 v1)
\end{lstlisting}

The number of \code{Frame} and \code{Body} constructors is linear in the number of active
fields. Their retained fields and search clauses occupy
$O(\sum_c(1+a_c)^2)$ target syntax nodes. The ordinary family backend also
specialises reconstruction helpers to each of the $m$ root sorts; its total
output is therefore bounded by $O(m\sum_c(1+a_c)^2)$ nodes. This duplication changes code size but not transition counts.

\section{Correctness by construction}
\label{sec:correctness}
The generated zipper operations are correct by construction: their clauses
implement the ordered traversal of Section~\ref{sec:overview}. We prove this
by relating the clauses of Section~\ref{sec:transitions} to a reference
machine that can suspend at an element. Its stack
$\kappa$ is a finite list of records $(c,\ell,\widehat{\mathbf v}_\ell)$:
returning a child fills the omitted field and resumes after $\ell$. Stacks are
sorted like trails: if $c$ is a constructor of sort $p$ and $E_{c,\ell}=Y_{p'}$,
pushing $(c,\ell,\widehat{\mathbf v}_\ell)$ turns $\kappa:\mathsf{Stack}(r,p)$
into a stack in $\mathsf{Stack}(r,p')$. The empty stack
$[\,]:\mathsf{Stack}(r,r)$ marks the root, and in each configuration below
the value $t$ or constructor $c$ belongs to the current sort of $\kappa$.

There are three running configurations:
\[
 \mathsf{Enter}_d(t,\kappa),\quad
 \mathsf{Return}_d(t,\kappa),\quad
 \mathsf{Advance}_d(c,\mathbf v,\ell,\kappa).
\]
There are also suspended configurations
$\mathsf{Yield}(c,\mathbf v,\ell,\kappa)$ and a terminal configuration
$\mathsf{End}$. Define $\scan_d(c,\mathbf v,\ell,\kappa)$ by selecting the
next active field $\ell'$ strictly after $\ell$ in direction $d$:
\[
\begin{array}{ll}
 E_{c,\ell'}=X:
   &\mathsf{Yield}(c,\mathbf v,\ell',\kappa),\\
 E_{c,\ell'}=Y_{p'}:
   &\mathsf{Enter}_d(v_{\ell'},(c,\ell',\widehat{\mathbf v}_{\ell'})::\kappa),\\
 \text{no }\ell':
   &\mathsf{Return}_d(c(\mathbf v),\kappa).
\end{array}
\]
Computing $\scan_d$ inspects one constructor description at the meta level
and costs no transition. The running configurations step as follows:
\begin{align*}
 \mathsf{Enter}_d(c(\mathbf v),\kappa)
   &\longrightarrow\scan_d(c,\mathbf v,e_d(c),\kappa),\\
 \mathsf{Return}_d(t,(c,\ell,\widehat{\mathbf v}_\ell)::\kappa)
   &\longrightarrow\scan_d(c,\widehat{\mathbf v}_\ell[\ell:=t],\ell,\kappa),\\
 \mathsf{Return}_d(t,[\,])&\longrightarrow\mathsf{End},\\
 \mathsf{Advance}_d(c,\mathbf v,\ell,\kappa)
   &\longrightarrow\scan_d(c,\mathbf v,\ell,\kappa).
\end{align*}
Here $e_+(c)=-1$ and $e_-(c)=a_c$ are the entry positions, and
$\widehat{\mathbf v}_\ell[\ell:=t]$ inserts $t$ at position $\ell$. Requesting
another element resumes a yield as an $\mathsf{Advance}$ configuration;
changing direction changes $d$ in that configuration. Each displayed
transition costs one abstract step.

The encoding of reference configurations as generated terms mirrors the traversal state: a reference stack becomes the
backward trail, a return record becomes $k_{c,\ell}$, and a yielded local position
becomes $(b_{c,\ell}(\widehat{\mathbf v}_\ell),v_\ell)$. Formally, stacks encode
as trails, $\lceil[\,]\rceil=\code{BNil}$ and
$\lceil(c,\ell,\widehat{\mathbf v}_\ell)::\kappa\rceil
=\mathsf{Snoc}(\lceil\kappa\rceil,k_{c,\ell}(\widehat{\mathbf v}_\ell))$, so
$\lceil\kappa\rceil:\mathsf{Trail}(r,p)$ for $\kappa:\mathsf{Stack}(r,p)$.

The relation $\approx$ relates $\mathsf{Enter}_d(t,\kappa)$ and
$\mathsf{Return}_d(t,\kappa)$ to $\seek_d(\lceil\kappa\rceil,t)$ and
$\rise_d(\lceil\kappa\rceil,t)$, respectively. A yield
$\mathsf{Yield}(c,\mathbf v,\ell,\kappa)$ relates to
$\code{Just}(\mathsf{Cursor}(\lceil\kappa\rceil,$ $b_{c,\ell}(\widehat{\mathbf v}_\ell),v_\ell))$.
An $\mathsf{Advance}_d$ of that yield relates to a move in direction $d$
from its cursor, and $\mathsf{End}$ relates to \code{Nothing}. Generated
records are constructed explicitly from field references, so inspecting
their outer tags terminates without forcing a source constructor.

\begin{theorem}[Correctness of the generated operations]
\label{thm:correctness}
Fix an accepted family description. (1)~Every generated search clause executes
one step of the reference machine and demands the same source constructor, if
any. (2)~Every cursor obtained by entry, \code{next}, \code{previous}, and
\code{replace} encodes a yield of the reference machine, and \code{close}
reconstructs its root with all replacements. (3)~On a finite, fully defined
input, entry and moves select the occurrences of the ordered traversal,
opposite moves undo each other, and replacement changes only the selected
occurrence.
\end{theorem}
The proof rests on five lemmas: simulation, mixed-operation preservation,
navigation, reversibility, and replacement.
Appendix~\ref{app:proof-thm-simulation} states them and sketches their proofs. Lean~4
mechanises simulation for finite values of single self-recursive datatypes, in
both directions (Section~\ref{sec:applications}); the other lemmas, and
simulation for families, are proved on paper.

A cursor is \emph{valid} when entry followed by navigation and replacement
produces it; cursor constructors are private (Section~\ref{sec:overview}).

\section{Cursor contracts and structural work}
\label{sec:cost}
An operation contract states what an operation returns, how many control
steps it takes, and which source constructors it forces, under stated
preconditions. A \emph{control step} is one reference transition of
Section~\ref{sec:correctness}: entering a constructor, resuming after a field,
or returning through a frame. \emph{Source demand} is the ordered list of
source constructors the operation forces; only entry steps force one. The
results are those of Theorem~\ref{thm:correctness}; the contracts add the step
count and the demand order. Demand order matters for partial and infinite
inputs, where forcing a subtree that the observation does not need can make it
diverge.

The cursor operations have the following contracts. Here $h$ is trail depth,
$q$ is the length of the operation's reference search, and $N,M$ count
constructor and element occurrences in the root.
\[
\begin{array}{l|c|l}
 \text{operation}&\text{control steps}&\text{source demand}\\ \hline
 \code{focus},\ \code{replace}&0&\text{none}\\
 \code{close}&h+1&\text{none}\\
 \code{openFirst}/\code{openLast}&q&\text{ordered reference search}\\
 \code{next}/\code{previous}&q&\text{ordered reference search}\\
 \code{traverse}&2N+M&\text{complete forward traversal}
\end{array}
\]
Entry and moves search to the next yield or boundary. The count for
\code{traverse} includes entry through \code{openFirst} and the final request
that returns the boundary. An ordered reference search forces the constructors
that the reference run matches, in the order it matches them. The table does
not bound $q$; for certified directions, Section~\ref{sec:infinite} bounds $q$
and the constructor demands, with uniform move bound $q\leq b_d(h)$
(Equation~\eqref{eq:family-bound}).

The contracts have three preconditions. Operations on an existing cursor
require a valid cursor (Section~\ref{sec:correctness}). The traversal count
requires a \emph{finite total root}: a finite, fully defined input. Certified
termination of entry or a move also requires \emph{productive demanded roots}:
each requested source value exposes its outer constructor in finite time. The
generated code does not check these preconditions, and no transition bound
helps when a demanded source constructor is undefined.

\kamot{} emits each contract in a JSON report;
Appendix~\ref{app:contract-semantics} gives the work grammar, the trace
descriptors, and the contract judgement.

\begin{theorem}[Entry contract]
\label{thm:entry-contract}
Fix an accepted family, a direction $d$, and a root value $t$, and let $R$ be
the reference run from $\mathsf{Enter}_d(t,[\,])$ to its first
$\mathsf{Yield}$ or $\mathsf{End}$. Then \code{openFirst} ($d=+$) or
\code{openLast} ($d=-$) on $t$ executes one generated search clause per step of
$R$, with the same source demand, on every finite prefix of $R$. If $R$ ends
after $q$ steps, the operation takes $q$ control steps and returns a cursor
that encodes the final $\mathsf{Yield}$, or \code{Nothing} at $\mathsf{End}$.
\end{theorem}

\begin{theorem}[Move contract]
\label{thm:move-contract}
Fix an accepted family, a direction $d$, and a valid cursor $z$, and let $R$
be the reference run from $\mathsf{Advance}_d$ of the yield that $z$ encodes
to the next $\mathsf{Yield}$ or $\mathsf{End}$. Then \code{next} ($d=+$) or
\code{previous} ($d=-$) from $z$ satisfies the conclusions of
Theorem~\ref{thm:entry-contract} for $R$.
\end{theorem}

\begin{theorem}[Local-operation contract]
\label{thm:cursor-contract}
At a valid cursor of depth $h$, \code{focus}, \code{replace}, and
\code{close} take $0$, $0$, and $h+1$ control steps, respectively, without
source demand.
\end{theorem}
\begin{proof}[Proof outline]
Theorem~\ref{thm:correctness} gives one clause per reference step with the
same demand and makes every valid cursor encode a yield; the local operations
satisfy their direct counts (Appendix~\ref{app:contract-semantics}).
\end{proof}

The structural cost model charges for inspecting an available constructor,
allocating a constructor, and reading or writing a field reference. Fields
are shared; copying a reference does not traverse its referent. Let
$a\geq1$ bound all source constructor arities. For a fixed description
$\boldsymbol\Phi=(\Phi_1,\ldots,\Phi_m)$, tag dispatch and constructor arities
are bounded, so each search clause has bounded structural cost independent
of input size and depth. We write $O_{\boldsymbol\Phi}$ when the hidden
constant may depend on $\boldsymbol\Phi$.

\begin{theorem}[Structural work]
\label{thm:cost}
A search of $q$ reference steps executes exactly $q$ generated search
clauses and incurs $O_{\boldsymbol\Phi}(q+1)$ structural work and allocation.
A complete traversal in one direction of a finite input, including the final
boundary result, executes exactly $q$ generated search clauses:
\begin{equation}\label{eq:cost}
 q=2N+M,
\end{equation}
where $N$ counts source constructor occurrences across the family's
unfolding and $M$ counts element occurrences. Its structural work is
$O_{\boldsymbol\Phi}(N+1)$. At a cursor of depth $h$, \code{focus} and \code{replace}
have constant structural overhead, and \code{close} takes
$O_{\boldsymbol\Phi}(h+1)$ work.
\end{theorem}
\begin{proof}
The first statement follows from part~1 of Theorem~\ref{thm:correctness}. One
emitted clause matches a bounded number of tags and allocates a bounded
number of records, each containing at most $a$ field references. The
backward trail requires one exposed-end update, never a scan of the
history. Taking the largest clause cost gives the constant.

In a complete traversal, every constructor is entered once and returned
from once, including the root's final return. These contribute $N$ entry
and $N$ return clauses. Each element yields once and is advanced past
once, contributing $M$ move clauses. No other search clause is executed.
This proves Equation~\eqref{eq:cost}, in either direction. Since
$M\leq aN$, the structural bound follows for fixed $\boldsymbol\Phi$.

Reading or replacing the focus accesses a fixed cursor field. Closing
reconstructs the terminal constructor and then removes one frame for
each of the $h$ ancestors, doing bounded work per frame; the final
\code{BNil} case returns the rebuilt root and is not counted as a step. No off-path
constructor is visited.
\end{proof}

The traversal count assumes no intervening calls to \code{close}.
Element replacement preserves the shape and hence the
count. Sharing in the source value is counted by unfolding: a shared
subtree reached in two distinct positions is traversed twice. Newly
allocated cursor records occupy $O_{\boldsymbol\Phi}(h+1)$ space at a suspended
position; the data they reference can be arbitrarily large. Retaining
old cursor versions can also retain their referenced data.

A single move need not cost $O(h+1)$. For
\begin{lstlisting}
data Sparse a = Empty | Item a | Split (Sparse a) (Sparse a)
\end{lstlisting}
a move between two elements can traverse an arbitrarily large intervening
empty subtree. Section~\ref{sec:infinite} gives a condition on the datatype
under which a function of $h$ bounds every move.

The structural model does not count the evaluation of source thunks or
elements, garbage collection, or client work.

\section{Compositional contracts for finite observations}
\label{sec:infinite}
We infer bounds on the transitions and constructor demands needed to reach
the next yield or boundary, without inspecting the input. The analysis
composes constructor summaries across sorts and then along the finite trail.

For a coinductive input $T^\infty=\nu Y.\,\Phi(X,Y)$, a reachable cursor still
has a finite history. Its finite-path contexts have the form
\[
 C_\infty=\mu Q.\,(B_\infty+K_\infty\times Q)
       \cong\bwd{K_\infty}\times B_\infty,
\]
where $K_\infty$ and $B_\infty$ are the recursive and element derivatives
evaluated at $T^\infty$. Stored siblings may be infinite; the trail spine
remains finite~\cite{Abbott2005}. For families, these contexts are indexed by
sort as in Section~\ref{sec:family-frontend}. Enumeration is \emph{responsive}
if each request terminates with a cursor or boundary.

\begin{corollary}[Finite observations]
\label{thm:productivity}
If demanded roots are productive, every finite reference observation is
returned by the generated search with the same source demands. Thus generated
cursor enumeration is responsive whenever every requested reference
observation is finite.
\end{corollary}
\begin{proof}
Theorems~\ref{thm:entry-contract} and~\ref{thm:move-contract} give the same
finite sequence of steps and demands. Each demand terminates, and navigation inspects no element or
constant field.
\end{proof}
An infinite label-free branch can still prevent a search
from reaching a sibling; the analysis must bound the search itself.

\subsection{Composing prefixes across sorts}
\label{sec:prefix-composition}
We first bound, for each sort, the search that enters one of its values, by
composing constructor summaries across the sorts it may reach before a yield.
Composition is already necessary for the lowered rose tree:
\code{List0Cons} carries no label, but its first child, a \code{Tree}, yields
immediately. The datatype dependency graph is cyclic, while the dependencies
followed \emph{before a yield} are acyclic.

A sort is \emph{nullable} when some finite traversal of a value of that sort
returns without yielding. Nullability $\mathsf{Null}$ and prefix edges $i\to_d j$ in direction
$d$ are the least relations closed under
\[
\frac{c\in\mathcal C_i\quad
      \text{$c$ has no $X$ field}\quad
      \forall \ell.\,E_{c,\ell}=Y_j\Rightarrow\mathsf{Null}(j)}
     {\mathsf{Null}(i)}
\]
\[
\frac{c\in\mathcal C_i\quad E_{c,\ell}=Y_j\quad
      \forall \ell'\prec_d \ell\ \text{active}.\;
      E_{c,\ell'}=Y_{j'}\wedge\mathsf{Null}(j')}
     {i\to_d j}.
\]
Here $\ell'\prec_d\ell$ means that field $\ell'$ precedes $\ell$ in direction
$d$. Prefix edges pass earlier nullable children because those children may
return empty.

A summary records separate \emph{yield} and \emph{return} outcomes, each with
a pair $(w,n)$ bounding transitions and demanded constructors. Impossible
outcomes are omitted. The entry summary $\mathcal E_j$ covers a search that
enters a value of sort $j$, up to its first yield or up to, but excluding,
its return; the continuation it returns to pays for that return. We evaluate
entry summaries in dependency order wherever the reachable prefix graph is
acyclic; an unresolved cycle gives \emph{unknown}.

Let $S(u)$ summarise a field suffix $u$, excluding its initial control action.
An empty suffix returns with $(0,0)$, an element yields with $(0,0)$, and
constants are skipped. For a recursive field $Y_j$,
\[
\begin{aligned}
 S(Y_j u).\mathsf{yield}
   &=\max\bigl(\mathcal E_j.\mathsf{yield},\;
        \mathcal E_j.\mathsf{return}+(1,0)+S(u).\mathsf{yield}\bigr),\\
 S(Y_j u).\mathsf{return}
   &=\mathcal E_j.\mathsf{return}+(1,0)+S(u).\mathsf{return}.
\end{aligned}
\]
Addition and maximum are componentwise. An alternative containing an
impossible outcome is omitted before its other terms are evaluated; an
outcome with no remaining alternatives is impossible. The $(1,0)$ charge
pays for the child's return. Entry adds $(1,1)$ for matching the constructor
and takes the maximum over constructors. A \emph{terminal continuation}
resumes the fields after a yielded element, and a \emph{frame continuation}
resumes the fields after a returning child; each adds $(1,0)$ for that
resuming transition to its suffix summary. Demands count occurrences, not distinct
heap objects.

Entry alone does not certify a move: a constructor with a label followed by
an endless label-free child yields on entry but diverges on resumption.
We therefore analyse every terminal and frame continuation. If all are
bounded, let $(w_B,n_B)$ and $(w_K,n_K)$ be the componentwise maxima over the
possible outcomes of terminal and of frame continuations, using $(0,0)$ for
an empty collection.
The move certificate at depth $h$ is
\begin{equation}\label{eq:family-bound}
 q\leq b_d(h)=w_B+h\,w_K+1,
 \qquad \#\mathsf{demands}\leq n_B+h\,n_K.
\end{equation}
The final $1$ covers a possible return from the root. The report also retains
the individual sort and suffix summaries.

\begin{theorem}[Compositional family contracts]\label{thm:family-prefix}
Fix an accepted family and direction $d$, with productive demanded roots.
A bounded root-entry summary $\mathcal E_r$ guarantees a yield or boundary
within
$\max(\mathcal E_r.\mathsf{yield},\mathcal E_r.\mathsf{return}+(1,0))$
transitions and constructor demands. An emitted move certificate guarantees
a yield or boundary within the bounds of Equation~\eqref{eq:family-bound}
from every valid cursor at depth $h$. Both guarantees hold for generated
searches on finite or infinite values, including auxiliary sorts.
\end{theorem}
\begin{proof}[Proof outline]
Induct over the acyclic prefix order and each constructor's field suffixes.
Every child that returns empty witnesses nullability, so prefix edges cover
all children a search can enter before yielding. Summing the terminal
segment, at most $h$ frame segments, and the root return gives the move bound.
Appendix~\ref{app:proof-family-prefix} gives the full proof.
\end{proof}

For the rose tree, \code{List0} is nullable and \code{Tree} is not. The forward
prefix graph has only \code{List0 -> Tree}, with entry summaries
\[
\begin{array}{c|cc}
 &\mathsf{yield}&\mathsf{return}\\ \hline
 \code{Tree}&(1,1)&\text{impossible}\\
 \code{List0}&(2,2)&(1,1)
\end{array}
\]
Both terminal and frame continuation maxima are $(3,2)$, giving $4+3h$
transitions and $2+2h$ demands per move, where $h$ counts lowered frames,
including list frames. Backward traversal has a prefix cycle and remains
uncertified. Adding \code{Silent} also makes forward traversal uncertified:
arbitrarily many label-free nodes may precede the next label. For one-sort
datatypes, a syntactic check suffices: in the direction of the move, every
recursive constructor has an element field before its first recursive field
(Appendix~\ref{app:local-guard}).

\paragraph{Charging each frame its own outcome.}
Equation~\eqref{eq:family-bound} charges every frame the componentwise maximum
over all frame outcomes, although only the last segment of a move can yield.
The report keeps each terminal and frame summary, so a client that
knows the trail can use a sharper bound.

\begin{theorem}[Trail-sum bound]\label{thm:trail-sum}
Under the assumptions of Theorem~\ref{thm:family-prefix}, let a valid cursor
have terminal record kind $\tau$ and trail frames $f_1,\ldots,f_h$, nearest
first, let $\mathcal B_\tau$ and $\mathcal K_f$ be the summaries of their
continuations, and suppose these summaries are bounded, as they are when a
move certificate is emitted. A move
reaches a yield or the boundary within the componentwise maximum of
$\mathcal B_\tau.\mathsf{yield}$, of
$\mathcal B_\tau.\mathsf{return}+\sum_{i<m}\mathcal K_{f_i}.\mathsf{return}+\mathcal K_{f_m}.\mathsf{yield}$
for $1\leq m\leq h$, and of
$\mathcal B_\tau.\mathsf{return}+\sum_{i\leq h}\mathcal K_{f_i}.\mathsf{return}+(1,0)$, omitting
terms with an impossible outcome.
\end{theorem}
\begin{proof}
As in the proof of Theorem~\ref{thm:family-prefix},
Lemma~\ref{lem:suffix-summary} bounds the terminal segment by
$\mathcal B_\tau$ and the $i$th frame segment by $\mathcal K_{f_i}$.
All earlier segments return; only the last can yield.
\end{proof}
Let $\mathcal B^{y},\mathcal B^{r},\mathcal K^{y},\mathcal K^{r}$ be the componentwise maxima of the yield and
return outcomes of terminal and frame continuations, taking $(0,0)$ when
no continuation has that outcome. For $h\geq1$,
Theorem~\ref{thm:trail-sum} gives the uniform bound
\begin{equation}\label{eq:separated}
 \max\bigl(\mathcal B^{y},\;\mathcal B^{r}+(h-1)\mathcal K^{r}+\mathcal K^{y},\;\mathcal B^{r}+h\,\mathcal K^{r}+(1,0)\bigr).
\end{equation}
For the rose tree, $\mathcal B^{y}=\mathcal K^{y}=(3,2)$ and $\mathcal B^{r}=\mathcal K^{r}=(3,1)$, so the demand
bound falls from $2+2h$ to $2+h$ and the transition bound stays $4+3h$.
For a label in a value of sort \code{Tree}, let $\delta$ count its ancestors
and $s$ the earlier siblings along its path, so $h=2\delta+s$.
Theorem~\ref{thm:trail-sum} gives $4+4\delta+s$ transitions and $2+\delta$
demands (Appendix~\ref{app:rose-depth}). Only the demand bound depends solely
on source depth. At $\nd{G}$, $\delta=2$, $s=1$, and $h=5$.
Section~\ref{sec:precision} checks these bounds on generated values.

\paragraph{Completeness.}
Call a family \emph{inhabited} when every sort occurs in some finite value of
a root sort and has a finite value that contains an element, and every
constant field type has a value.

\begin{theorem}[Completeness]\label{thm:completeness}
Fix an inhabited accepted family and a direction $d$.
\begin{enumerate}
\item For each root sort $r$, $\mathcal E_r$ is bounded if and only if one
constant bounds the transitions of root entry on all finite values of sort
$r$.
\item A move certificate is emitted if and only if some function of $h$ alone
bounds the transitions of every move from a valid cursor at depth $h$ on
finite values.
\item If no move certificate is emitted, some infinite value whose
constructors are all defined has a valid cursor whose move diverges.
\end{enumerate}
Parts 1 and 2 also hold with constructor demands in place of transitions.
\end{theorem}
\begin{proof}[Proof outline]
Soundness is Theorem~\ref{thm:family-prefix}. Repeating a reachable prefix
cycle gives unbounded finite entry costs. Inhabitation places each unknown
continuation after a valid cursor at fixed depth; repeating its cycle then
gives unbounded move costs, and following it forever gives a productive
input with a divergent move. Appendix~\ref{app:proof-completeness} gives
the constructions.
\end{proof}
For an inhabited family, the moves in each direction are therefore either
bounded by an affine function of $h$ or bounded by no function of $h$.
Completeness is relative to \kamot{}'s traversal and lowering: a loopless
traversal~\cite{Bird2006Loopless} or a snoc lowering of lists can bound moves
that this one cannot. The proof
uses inhabitation: a frame whose child sort holds no element in any finite
value cannot occur in a valid cursor's trail on a finite value, yet its
continuation is analysed and can be reported unknown.

\subsection{Using the bounds}
The resumable backend exposes each reference transition as \code{step},
retaining its control position, typed trail, and strict cached depth. The driver
\code{runFor} with budget $\beta$ executes at most $\max(0,\beta)$ transitions
and returns a result or suspended state. For $\beta>0$, a scheduler runs
\code{next} directly when $b_d(h)\leq\beta$; otherwise, or without a
certificate, it runs \code{runFor}~$\beta$.

\begin{corollary}[Contract-directed slicing]\label{cor:scheduling}
For a move from a valid cursor on productive demanded roots, each scheduler
invocation with budget $\beta>0$ executes at most $\beta$ reference transitions.
Positive-budget resumptions
preserve the unbounded search's results and ordered demand trace.
\end{corollary}
\begin{proof}
Theorem~\ref{thm:family-prefix} bounds the direct case; the driver's counter
bounds each slice. Suspension preserves the search state.
\end{proof}
Appendix~\ref{app:scheduling} measures the scheduling policy.

\paragraph{Admitting a complete observation.}
A client can reserve both a work budget $\beta$ and a demand budget $\chi$.
It admits a move when a certificate is present and
\[
 b_d(h)\leq\beta,\qquad n_B+h\,n_K\leq\chi.
\]
The test reads only the certificate and cached depth. An admitted move
completes within both budgets on productive input. Rejection demands no
source but may defer a move whose actual cost fits the budgets. The dynamically budgeted
alternative in Section~\ref{sec:admission-eval} spends demand credit only on
entry, allowing returns and advances without forcing source constructors.

If a terminal suffix has no return outcome, the search must yield before
consulting the trail. Its local work and demand bounds then suffice, even at
large depth. Inspecting the terminal tag does not force its saved subtrees.
Appendix~\ref{app:admission} evaluates this refinement.

\section{Resumable eager folds}
\label{sec:folds}
A fold replaces completed subtrees by their results. Following McBride's
dissection~\cite{McBride2008}, its frames distinguish completed results $J$
from pending source subtrees $T$. For the rose tree, the container-level
frame is
\[
 \mathsf{Frame}_{\mathrm{fold}}(X,J,T)
   =X\times\bwd{J}\hole\fwd{T}.
\]
Completed siblings lie to the left of the hole and pending siblings to its
right; setting $J=T$ recovers the ordinary frame. After any finite execution
prefix, the completed spine is finite even when pending subtrees are infinite.

\kamot{} implements this distinction after container lowering, with indexed
frames and one result type $J$ shared by all sorts. The algebra for sort $i$
has type $\Phi_i(X,J,\ldots,J)\to J$. Appendix~\ref{app:folds} gives the
construction and discusses extending it to separate result types per sort.

The reference fold exposes a constructor, evaluates its recursive children
left to right, and applies the algebra to the completed layer. Child and
algebra results are forced to weak head normal form (the outer constructor or
function closure); element and constant fields remain lazy unless demanded
by the algebra. The machine therefore enters even children ignored by the
algebra.

The interface provides \code{start}, \code{step}, \code{pending},
\code{runFor}, and \code{run}. The \code{step} clauses are nonrecursive; each
transition descends, applies the algebra, or returns. A completed fold over $N$ constructors
therefore takes exactly $3N$ transitions, one of each kind per constructor,
and agrees with the reference eager fold (Theorem~\ref{thm:fold}). Fuelled
execution partitions this sequence into resumable slices. Source forcing
and algebra work lie outside the structural count, so fuel cannot interrupt a
divergent thunk or callback. On infinite input, each fuelled slice still
suspends after finitely many transitions if its demands and algebra
applications terminate, although the fold never finishes.

Theorem~\ref{thm:end-to-end} (Appendix~\ref{app:folds}) covers the cursor and
eager-fold backends: compilation terminates, the generated types follow the
indexed schemas, and each backend satisfies its stated contracts.

\paragraph{Complete batch plans.}\label{sec:selection}
Cursor enumeration and an eager fold can both collect the elements of a
finite total root in field order. Appendix~\ref{app:selection} gives a
selector based on their $2N+M$ and $3N$ control counts, and
Appendix~\ref{sec:batch-crossover} measures its choices. These counts exclude
source forcing and output construction and do not predict elapsed time.

\section{Evaluation}
\label{sec:applications}
\label{sec:prototype}
\kamot{} implements the family IR and declaration frontend of
Section~\ref{sec:source} in OCaml. It emits JSON contracts and Haskell
code that uses indexed GADTs without casts or reflection. After validating the
generated code and certificates, we test four claims:
\begin{itemize}
\item[(C1)] Every tested direction without a move certificate has a checked
productive input with a divergent move. Both emitted and refined bounds hold
on generated values (Section~\ref{sec:precision}).
\item[(C2)] Admission from emitted bounds answers every local query, as
fuelled search does; on long returns it avoids discarded evaluations but can
lose answers (Section~\ref{sec:admission-eval}).
\item[(C3)] Generated cursors match the handwritten indexed stack's allocation,
with overlapping timing ranges, and are several times faster than generic
zippers (Section~\ref{sec:runtime-eval}).
\item[(C4)] For families of up to 32 sorts and 64 constructors, fresh
\code{-O2} builds take 0.16--2.76\,s for cursors and 0.27--3.46\,s for folds;
the largest cursor module is 222\,KiB (Appendix~\ref{app:family-scaling}).
\end{itemize}
The generator's instrumented one-sort variant counts entry, return, and
advance clauses. Three independent tools check the output. A Python validator checks the work ASTs and
composition metadata in the reports. A certificate checker reads the source
family and prefix certificate, without importing the compiler or inference
code, and checks nullability, acyclicity of prefix dependencies, local bound
inequalities, and composition along the trail. A source interpreter compares
execution traces and results with generated Haskell, including on partial and
infinite-prefix inputs. The suite checks 81 source families and rejects five
deliberately corrupted certificates. On nine families, 201 comparisons
check that generated resumable Haskell and the source interpreter agree on
traces and results.

The artifact's Lean~4 development proves the finite one-sort results in both
directions: lockstep simulation for a generated dispatch table, with the same
source constructor matched at each step (part~1 of Theorem~\ref{thm:correctness}), termination, root-entry and move bounds
(Theorem~\ref{thm:family-prefix}), trail-sum bounds
(Theorem~\ref{thm:trail-sum}, under a move certificate), and completeness for transitions and demands
(parts~1 and~2 of Theorem~\ref{thm:completeness}).

The completeness witnesses are reached by root entry and finitely many
completed moves. A separate forward model of constructor unfoldings proves
part~3: a productive root leads to a cursor whose next move diverges.
This model retains saved fields and reconstructed return values. The finite
backward results use mirroring and a Lean formulation of the compiler's backward
analysis. Families, lowering, infinite-input bounds, and the link to printed
Haskell remain unmechanised. For one sort, the prefix analysis reduces to a
syntactic guard. The development uses only core Lean and its
standard axioms.

Source fields in generated records remain lazy; cached depth and fuel counters
are strict. The 201 trace comparisons pass under both \code{runghc} and
\code{ghc -O2}, including partial inputs with undefined unvisited fields.
Across 108 configurations, the \code{-O2} admission audit checks that producer
invocations equal recorded Enter demands. Separate tests show that zero
demand credit leaves undefined children unforced, while zero fuel leaves
the search state unforced. These tests observe logical
traces, counts, and partial-input behaviour, not actual thunk-forcing order
in optimised code. Part~1 of Theorem~\ref{thm:correctness} relates each
generated pattern match to a reference demand; that compiled code forces
source constructors in this order is tested, not proved.

The runtime harness compares \kamot{} cursors with a handwritten explicit stack, direct
recursive rescan, and two generic baselines: \code{zippers-0.3.2}'s generic
\emph{Jacket}/\emph{Path}~\cite{KmettZippers}, and a GHC~9.12 port of
McBride's \code{Diss} zipper~\cite{McBride2008}. It records elapsed time,
runtime system (RTS) allocation, and post-collection live bytes independently
of clause counts. The cursor, scheduling, and batch-crossover measurements
use GHC~9.12.1 at \code{-O2} on an Apple M3 Max running macOS~26.5.2,
with \code{+RTS -T -A32m}; the one-sort cursor-comparison and fold builds also use
\code{-threaded}. The artifact retains raw measurements, generated
sources, and build metadata. We report medians and ranges of seven
fresh-process samples (three for compiler builds), treating overlapping
ranges as unresolved. Appendix~\ref{app:cursor-evaluation} gives detailed
protocols and validation coverage.

\subsection{Certificate precision}
\label{sec:precision}
We analyse a corpus of 19 datatype models, each with a forward and a backward direction, and the trace suite's 81 generated families.
The analysis certifies 10 of 38 corpus directions. All 28 unknown corpus
directions, all 82 unknown generated directions, and every unknown root
entry have independently checked divergence witnesses
(Appendix~\ref{app:counts}). Truncating an entry witness gives unbounded
finite entry costs; for a move witness, costs grow while cursor depth stays fixed.

For each certified direction, an independent interpreter checks every move
on 200 generated finite values. No move exceeds its bound, and 7 of the 10
emitted transition bounds are attained exactly (Table~\ref{tab:precision}).
Only boundary moves, which return through every frame, attain them; moves
that yield reach at most 0.86 of the bound (medians 0.29--0.67).
On all 3,847 moves, observed costs never exceed the trail-sum bound of
Theorem~\ref{thm:trail-sum}, which never exceeds the separated bound, which
never exceeds the emitted one; some move reaches its trail-sum bound in every
direction. For rose-tree values of sort \code{Tree}, the source-tree bounds
$4+4\delta+s$ and $2+\delta$ are attained and never exceeded. The values are
shallow ($h\leq7$), so these checks test the bounds, not their slopes. The divergence witnesses are checked independently of
Theorem~\ref{thm:completeness}, which guarantees that they exist for
inhabited families. The models admit unrestricted
values: the \code{Data.Map} shape, for example, omits balance invariants, so
precision on constrained library values is not established.

In the syntax models, annotation position explains the outcomes: annotations before children give
forward certificates, and annotations after children give backward ones.
Unannotated applications can delay the next variable indefinitely.
Repeated JSON nulls, DOM comments, YAML sequences, and empty trie nodes prevent bounds in either direction.

\begin{table}[htbp]
\centering
\caption{Certified directions in the 19-datatype corpus. Emitted gives the
transition and demand bounds of Equation~\eqref{eq:family-bound}; Separated
gives those of Equation~\eqref{eq:separated} for $h\geq1$. Moves is the total
over 200 generated values (empty values contribute none); Ratio is the
largest observed transition count divided by the emitted bound. The other 28
directions have divergence witnesses.}
\label{tab:precision}
\begin{tabular}{llllrr}
\hline
Datatype & Dir. & Emitted & Separated & Moves & Ratio\\
\hline
List & fwd & $4+h$; $1$ & $4+h$; $1$ & 90 & 1.00\\
List & bwd & $2+h$; $0$ & $2$; $0$ & 90 & 1.00\\
Event log & fwd & $4+h$; $1$ & $4+h$; $1$ & 112 & 1.00\\
Event log & bwd & $2+h$; $0$ & $2$; $0$ & 112 & 1.00\\
Rose tree & fwd & $4+3h$; $2+2h$ & $4+3h$; $2+h$ & 352 & 1.00\\
File system & fwd & $4+3h$; $2+2h$ & $4+3h$; $2+h$ & 246 & 1.00\\
\code{Data.Map} shape & fwd & $6+3h$; $2+h$ & $6+3h$; $2+h$ & 178 & 1.00\\
Located module & fwd & $5+3h$; $2+2h$ & $4+3h$; $2+h$ & 715 & 0.88\\
Annotated lambda & fwd & $3+2h$; $1+h$ & $2+h$; $1$ & 1,077 & 0.67\\
Postfix lambda & bwd & $3+2h$; $1+h$ & $2+h$; $1$ & 875 & 0.67\\
\hline
\end{tabular}
\end{table}

\subsection{Admission before source demand}
\label{sec:admission-eval}
As an application, a prefetching service uses the emitted bounds to decide
whether to start a query on the mutually recursive \code{Expr}/\code{Decl}
family before an edit may cancel it (Appendix~\ref{app:admission}). In local
workloads, certified admission answers all 256 queries, while deferring every
query until after the edit loses half. On long returns with eight demand
credits and 50\% cancellation, it avoids 1,024 discarded constructor
evaluations and returns the same 128 answers as fuelled search, but blanket
deferral does the same there; at larger budgets conservative bounds lose
13--15 answers. The study models a scheduling kernel, not editor
responsiveness.

\subsection{Comparison with handwritten and generic cursors}
\label{sec:runtime-eval}
An annotation-service session makes 2,048
successive requests on a depth-4,096 skewed syntax tree and edits every 64th
annotation. The \code{zippers} baseline takes an explicitly written preorder
traversal; McBride's dissection uses an ordered polynomial encoding and
generic zipper moves (Table~\ref{tab:runtime}).

\begin{table}[htbp]
\centering
\caption{Incremental annotation: median time and allocation per request;
live heap for one retained state. Code identifies the code written for each
datatype; Bound indicates an emitted move bound.}
\label{tab:runtime}
\begin{tabular}{llcrrr}
\hline
Implementation & Code & Bound & ns/request & bytes/request & live KiB\\
\hline
\kamot{} cursor & declaration & yes & 107 & 200 & 426\\
Handwritten explicit stack & cursor & no & 113 & 263 & 394\\
McBride generic dissection & encoding & no & 560 & 1,058 & 1,273\\
External \code{zippers} & traversal & no & 690 & 4,237 & 3,126\\
Direct recursive rescan & search & no & 4,940 & 19,045 & 394\\
\hline
\end{tabular}
\end{table}

Time includes navigation, edits, and reconstruction; the dissection port also
converts its polynomial result back to the native tree inside that interval.
The dissection port's navigation and edits alone take 270\,ns per request, still above the
generated cursor's complete-session time. The generated cursor takes similar time to the
handwritten stack and allocates fewer bytes in this workload. On a mutually
recursive syntax family, a separate comparison with an independently
handwritten, sort-indexed stack (Appendix~\ref{app:family-service}) gives identical per-request allocation and overlapping
timing ranges. An unindexed handwritten
stack is faster on a deep chain of lets (38 versus 58\,ns per request) but allocates
more (309 versus 253 bytes). Separate fresh
processes measure live heap after checking that all implementations produce equivalent outputs.
Appendix~\ref{app:cursor-evaluation} gives the dissection phase breakdown, the
family comparison, full protocols, the streaming-repair workload, and
retained-state details.

\section{Related work}
\label{sec:discussion}
\paragraph{Derivatives, containers, and dissections.}
Huet's zipper~\cite{Huet1997} pairs a focus with a one-hole context;
McBride~\cite{McBride2001} calculates these contexts for regular types.
Container derivatives cover inductive and coinductive fixed points,
including finite paths into potentially infinite values~\cite{Abbott2005};
indexed containers give normal forms for strictly positive
families~\cite{Altenkirch2015}.
Our closest predecessor, dissection~\cite{McBride2008}, represents suspended
traversals with contexts whose processed and pending fields may have different
types. It derives generic navigation and tail-recursive catamorphisms; our
fold backend adopts its processed/pending distinction. \kamot{} specialises
these constructions to ordered families and derives contracts for their
generated operations.

\paragraph{Generic zippers.}
Generic Haskell derives zippers from fixed-point datatype
descriptions~\cite{HinzeJeuring2003}. The multirec zipper indexes its stack by
root and focus types for mutually recursive families~\cite{Rodriguez2009}.
Our indexed \code{Trail} GADT (Theorem~\ref{thm:family-trail}) specialises
this construction to generated first-order frames; its path
equation~\eqref{eq:family-path} uses the path interpretation of a linear
system's least solution due to Kleene and Conway~\cite{Conway1971}. Heterogeneous \code{Data}-based
zippers~\cite{Adams2010} support broader navigation; we exclude arbitrary
\code{Data}-instance graphs. Kiselyov derives a zipper from any traversal by
capturing its delimited
continuation~\cite{KiselyovShan2007,Kiselyov2005Zipper}; its cursor is the
suspended traversal, without a typed trail or cost bound.

\paragraph{Data-structure and resource-guided synthesis.}
Data representation synthesis compiles relational specifications through
user-chosen decompositions~\cite{Hawkins2011}; Cozy synthesises collections
from retrieval specifications~\cite{Loncaric2016}. Resource-guided synthesis
prunes type-directed search with user-supplied bounds, producing programs
correct by construction~\cite{Knoth2019}. These systems take operation
specifications or bounds as input. \kamot{} takes a datatype declaration,
fixes one representation, the derivative trail, and emits its bound.

\paragraph{Grammar analyses and productivity.}
With sorts as nonterminals and elements as terminals,
$\mathsf{Null}$ (Section~\ref{sec:prefix-composition}) is grammar nullability,
forward prefix edges form the left-corner relation, and cycles express left
recursion~\cite{Aho2007}, or right recursion in the backward direction.
Parsing expression grammars use a similar fixed-point check to reject left
recursion and ensure termination~\cite{Ford2004}. That check answers yes or
no. Our analysis separates yield from return, covers entry and every
resumption, and weights the relation with transition and demand bounds; these
bounds compose along the trail and transfer to generated code.

Productivity analyses establish that lazy definitions produce every finite
output prefix, using sufficient conditions on list equations~\cite{Sijtsma1989}, sized
types~\cite{Hughes1996,Abel2010}, or a decision procedure for pure stream
specifications~\cite{Endrullis2010}. We assume a productive source and ask
whether its traversal is productive for every input of a datatype. Stream
processors typed by a least fixed point nested in a greatest one take
finitely many input steps per output~\cite{Hancock2009}; a certified move is
such a processor with an explicit bound on its inductive part.
Constant-delay enumeration bounds time between answers after
preprocessing~\cite{Bagan2007,Segoufin2013}; our bounds need no preprocessing
but grow with trail depth.

\paragraph{Folds and first-order machines.}
Our fold language follows recursion schemes~\cite{MeijerFokkingaPaterson1991}.
First-order stacks descend from continuation-based
interpreters~\cite{Reynolds1972} and defunctionalised machines~\cite{Danvy2008}.
We apply this pattern to a restricted eager language: frame payloads come from
ordered derivative contexts, and contracts count structural transitions,
source forcing, and client algebra separately.

\section{Conclusion and Limitations}
\label{sec:limitations}
This paper showed that the derivative-based zipper of an ordered regular
datatype can be compiled together with a bound on the work and source demand
of each move. The analysis that produces these bounds rests on keeping the
cost of a search that yields separate from that of a search that returns.
With this separation, a search that must yield before it reaches a saved frame
is bounded independently of the trail, and summaries computed for different
sorts can be composed. Container lowering is what lets the same rules apply
to the rose tree, whose list cells carry no label of their own.

In our experience, whether a direction is certified depends more on the order
of fields than on the overall shape of the datatype. The rose tree, for
example, is certified forward but not backward, and for a single datatype
Appendix~\ref{app:local-guard} reduces the question to a syntactic test on
field order.

The evaluation also shows where the bounds are loose. Only boundary moves,
which return through every frame, reach the emitted transition bound, while
moves that yield stay below 0.86 of it. The trail-sum bound is considerably
closer, although \kamot{} does not yet emit it. In the admission study, the
bounds saved source evaluations when queries were cancelled during long
returns, but at larger budgets their conservatism cost 13--15 answers that
fuelled search delivered.

Several limitations remain. \kamot{} accepts only the regular, strictly
positive language of Section~\ref{sec:source}, and its folds use a single
result type for all sorts. We test, but do not prove, that compiled Haskell
forces source constructors in the reference order, and the Lean development
covers only single self-recursive datatypes; families, container lowering,
and bounds on infinite inputs are proved on paper. Finally, we have not yet
evaluated access patterns beyond those in Section~\ref{sec:applications}, or
integration into an interactive editor.

\section*{AI Use}
The authors used Claude Opus 5.5 (Anthropic, model
\texttt{claude-opus-5-5}), accessed through Claude Code, and GPT-6 Astra
(OpenAI, model \texttt{gpt-6-astra}), accessed through Codex, from July
through November 2026. The tools were used to revise and edit text; to
implement parts of the compiler prototype, the benchmarks, and the
validation checkers; and to draft proof arguments, the Lean development, and
tests. The system
design, the theorems, and the claims are the authors' own. The authors
reviewed all generated text, code, and proofs, checked the correctness of
every reported result, and take full responsibility for the content of
this paper. Reported measurements come from the benchmark and validation
programs in the artifact. The one-sort results listed in
Section~\ref{sec:applications} are mechanised in Lean~4; the other proofs are
not mechanised.

\section*{Data-Availability Statement}
The artifact is available at
\url{https://doi.org/10.5281/zenodo.24193628}. It contains \kamot{}, the
independent checkers, the benchmarks with raw results, and the Lean~4
development, which covers single self-recursive datatypes only. Timings are
machine-dependent. We will submit the artifact for Artifact Evaluation.

\bibliographystyle{ACM-Reference-Format}
\bibliography{compiler-references}
\appendix
\section{The local entry condition}\label{app:local-guard}
For a one-sort datatype, a local syntactic check suffices to certify that
moves in one direction are productive. Call
$\Phi$ \emph{entry-guarded in direction $d$} when every constructor
containing a recursive field has an element field before its first
recursive field in direction $d$. Equivalently, its first active field
is $X$. Nonrecursive constructors impose no condition. The check takes
one pass over the fields and is independent of the input value.

\begin{theorem}[A sufficient productivity certificate]
\label{thm:guard}
Suppose $\Phi$ is entry-guarded in direction $d$ and let $f\geq0$ bound its
number of recursive fields per constructor. On productive demanded roots, entry in direction $d$ yields or terminates in at most two search
clauses. A move in direction $d$ from a valid cursor at depth $h$ yields
or terminates within $q$ search clauses, where
\begin{equation}\label{eq:guard}
 q\leq g_d(h)=1+(2f+1)(h+1).
\end{equation}
In particular, repeated moves in that direction are
productive, even when off-path fields are infinite.
\end{theorem}
\begin{proof}
On entering a constructor with any recursive field, the guard makes the
first active field an element, so entry yields without descending. A
nonrecursive constructor either yields an element or has no active
fields and returns immediately. Root entry therefore costs one clause
to yield, or one entry and one return clause to terminate.

For a move, count its initial advance clause. Before its next yield,
every newly entered child is either an immediate yield or an empty
nonrecursive constructor; the latter costs one entry and one return
clause. Thus a search cannot build an unbounded chain of new frames
without yielding. It can finish only the current constructor and its
$h$ existing ancestors. At most $f$ previously unvisited recursive fields
per such constructor cost at most two clauses each, and completing each
of these $h+1$ constructors costs at most one further return clause.
Adding these costs gives Equation~\eqref{eq:guard}. The demanded roots
are available in finite time by assumption, so the finite clause bound
implies termination of the search.
\end{proof}

\section{Batch-plan selection details}
\label{app:selection}
The cursor and fold contracts expose different capabilities and control costs.
A client can choose a plan from these contracts by specifying the required
result and meeting the plan's preconditions.

The cursor and eager-fold batch plans are compared only for a common, explicitly declared
workload. For a finite total root $t$, let
$\mathsf{elements}_\Phi(t)$ be the direct elements of $t$ in the forward
field order of Section~\ref{sec:source}. A cursor client obtains this result
by recording every focus of a complete forward traversal. A fold client uses
the algebra whose direct-element field denotes a singleton sequence and whose
recursive fields denote their already collected sequences, concatenated in
field order. The implementation represents sequences as difference lists
(functions that prepend a sequence to a supplied tail); their composition
and final list construction are client work, outside the machine transition count.

\begin{lemma}[Common ordered-elements batch semantics]
\label{lem:batch-semantics}
For every finite total root $t$, complete cursor enumeration and the eager
fold with the ordered-elements algebra both return
$\mathsf{elements}_\Phi(t)$.
\end{lemma}
\begin{proof}
The cursor result follows from the traversal specification: every direct
element is yielded once, in constructor-field order. For the fold, induction
on $t$ shows that each completed recursive field denotes the elements of that
child. The layer algebra concatenates direct-field singletons and completed
child sequences in constructor-field order. Both computations
therefore produce the same sequence. The two machines need not have the same
intermediate demand order: the cursor yields during descent, whereas the
eager fold applies its algebra after recursive children have completed.
\end{proof}

A client profile records the requested result, whether it needs local
replacement or resumption, and whether forward observations require a
productivity certificate. Profiles describe one-shot observation, resumable
cursor interaction, or a finite batch; batch profiles request no cursor
operations. The artifact's policy client supports the batch
result \code{ordered-elements}. For this comparison, the profile supplies
exact constructor and element counts $N$ and $M$ (Equation~\eqref{eq:cost}), for example from a maintained index or earlier analysis.
The selector never traverses a root to obtain them.
It first eliminates plans that lack required capabilities. It chooses root
rescan for a one-shot observation and the persistent cursor for edits or
resumable productive observations. For the declared ordered-elements batch,
Lemma~\ref{lem:batch-semantics} justifies comparing $2N+M$ and $3N$: the
cursor is selected when $M<N$, and the eager fold otherwise.

\begin{theorem}[Contract-directed plan selection]
\label{thm:plan-selection}
For an accepted description and a profile satisfying the preconditions of
some plan with the requested capabilities, the selector returns a plan with every requested capability.
For an ordered-elements batch profile with supplied exact $N$ and $M$, its
selected plan has a reported control count no larger than that of the other batch plan.
\end{theorem}
\begin{proof}
The composition metadata is assembled from public operation clauses, so its
capabilities are exactly those supplied by the generated interface.
Lemma~\ref{lem:batch-semantics} gives the common result semantics.
Theorem~\ref{thm:cost} establishes $2N+M$ and
Theorem~\ref{thm:fold} establishes $3N$ for a completed eager fold.
Comparing the two is equivalent to comparing $M$ with $N$. The other cases
are capability exclusions: root rescan has no saved focus, while the fold
backend exposes neither cursor replacement nor cursor resumption.
\end{proof}

\subsection{Batch measurements}
\label{app:batch-measurements}
\label{sec:batch-crossover}
We test the selector on a
chain constructor with $\eta$ direct element fields, $\eta\in\{0,1,2,4,8\}$, and
4,096 links, so $N=4{,}097$ and $M=4{,}096\eta$. Every policy builds and consumes
the same ordered element list. Comparing $2N+M$ with $3N$ selects the faster of
the cursor and the unbounded eager fold at all five tested densities: the cursor is
2.9 times faster at $\eta=1$, and the fold 2.1 times faster at $\eta=2$. At $\eta=1$ the
counts differ by one transition, so this agreement does not show that the
count margin predicts the time margin. Table~\ref{tab:batch-crossover} reports the measurements.

Each chain constructor places its $\eta$ direct element fields before its
recursive child. The timed observer consumes
every cell and value of the output list, so fold algebra and output
construction are included. Exact list comparisons precede timing. Each of
seven fresh-process samples uses 64 preconstructed roots; roots with $\eta>0$ have distinct element seeds. Roots with $\eta=0$ may share, and serve only as an empty-output control.

\begin{table}[htbp]
\centering
\caption{Common-output batch traversal: median $\mu$s per tree. The count
selector compares the generated cursor with the unbounded eager fold. Fuelled
folds and direct recursion are additional baselines.}
\label{tab:batch-crossover}
\begin{tabular}{r l r r r r}
\hline
$\eta$ & Selected plan & Cursor & Fold & Fold, fuel 31 & Recursive\\
\hline
0 & Cursor & 20.2 & 40.2 & 54.9 & 5.3\\
1 & Cursor & 52.5 & 152.6 & 150.0 & 89.7\\
2 & Fold & 252.4 & 120.5 & 236.0 & 180.6\\
4 & Fold & 331.8 & 180.7 & 222.9 & 191.2\\
8 & Fold & 620.8 & 307.0 & 324.3 & 223.3\\
\hline
\end{tabular}
\end{table}

The unbounded fold is faster than the fuel-31 fold at every density except
$\eta=1$, where the medians differ by 2\% and the ranges overlap. Control counts alone do not determine elapsed
time: the fold's count is $3N$ at every density, while its time ranges from 40.2 to 307.0\,$\mu$s. Direct recursion is faster than both generated plans at $\eta=0$ and
$\eta=8$ but provides neither
generated cursor navigation nor the fold's suspension interface.

The batch timings depend on GHC inlining: at
$\eta=1$ GHC inlines \code{next} into the enumeration loop, which allocates
104 bytes per element, while at $\eta=2$ \code{next} remains a call and the
loop allocates 168. Allocation in the timed region then grows from 27\,MB, within one
\code{-A32m} nursery, to 88\,MB, so the cursor's 4.8-fold slowdown between
these densities reflects inlining and collection as well as element count.

\section{Eager-fold construction, proof, and measurements}
\label{app:folds}
The cursor backend preserves source subtrees because navigation can revisit
them. A fold instead replaces each completed child by a result. This change
affects the frame payloads and the continuation clauses, while leaving the
history representation intact. For one sort, an algebra
$\alpha:\Phi(X,J)\to J$ determines the catamorphism
$\mathsf{cata}\,\alpha:T\to J$: recursively fold each child to a $J$, then
apply $\alpha$ to that layer. For a family, the backend emits one indexed machine
interface per root sort and an algebra family
$\alpha_i:\Phi_i(X,J,\ldots,J)\to J$. \kamot{} uses a common result
type $J$ across sorts. Each algebra processes one completed constructor layer.

\subsection{Refining the recursive derivative}
For left-to-right evaluation, a frame for recursive field $\ell$ retains
earlier recursive fields at type $J$, later ones at type $T$, and
nonrecursive fields at their original types:
\begin{equation}\label{eq:fold-frame}
 K_{\mathrm{fold}}(X,J,T)=
 \sum_{c,\ell:E_{c,\ell}=Y}
 \left(\prod_{\ell'<\ell} E_{c,\ell'}[X,J]\right)
 \times
 \left(\prod_{\ell'>\ell} E_{c,\ell'}[X,T]\right).
\end{equation}
Here $E[X,Z]$ replaces the recursive variable by $Z$. Setting $J=T$
recovers the ordinary recursive derivative. Keeping the two arguments
distinct is the dissection refinement of
McBride~\cite{McBride2008}; deriving a tail-recursive catamorphism this
way is established prior work. Our backend specialises these records to
the input constructors, attaching a cost and demand contract to each transition.

For $\Phi(X,Y)=X+Y\times Y$, the two recursive positions give
$K_{\mathrm{fold}}=T+J$: the first frame retains the right subtree while
waiting for the left result, and the second retains the left result while
waiting for the right. The path--trail correspondence applies to
$\mu Q.\,(1+K_{\mathrm{fold}}\times Q)$, representing the suspended
ancestors by $\bwd{K_{\mathrm{fold}}}$ with the nearest frame exposed.
Replacing a frame after completing a child takes constant structural
work for fixed $\Phi$.

For a family, a frame from parent sort $i$ to child sort $j$ retains earlier
family children as results and later children as source values:
\[
 K^{\mathrm{fold}}_{ij}(X,J,\mathbf T)=
 \sum_{c\in\mathcal C_i,\,\ell:E_{c,\ell}=Y_j}
 \left(\prod_{\ell'<\ell}E_{c,\ell'}[X,J]\right)\times
 \left(\prod_{\ell'>\ell}E_{c,\ell'}[X,\mathbf T]\right).
\]
Here $E[X,J]$ replaces every recursive sort by the common result type.
The generated Haskell type indexes a frame by its parent and child sort and
indexes a trail by its root and current sorts. GADT indices enforce the
heterogeneous dissection invariant: a $Y_j$ return resumes only a frame
awaiting its result, at the statically known parent sort.

One extension would give each sort its own result type $J_i$, with
algebras $\alpha_i:\Phi_i(X,\mathbf J)\to J_i$. An earlier child of sort $j$
would then contribute a $J_j$, while a pending child would retain its source
type $T_j$. For the lowered rose tree, \code{Tree} could produce $J$ and
\code{List0} could produce $\fwd{J}$, making the container-level algebra
$X\times\fwd{J}\to J$ directly expressible. The current backend uses a common
$J$ for all sorts; this extension would require corresponding changes to
the generated layers, machine states, and correctness argument.

\subsection{Evaluation strategy}
The reference fold first exposes the current source constructor. It then
recursively evaluates its $Y$ fields in declaration order, forcing each
child result to weak head normal form. Finally it applies $\alpha$ to the
completed layer and forces its value to weak head normal form.
Element and constant fields are carried unchanged; the algebra may demand
them. Weak head normal form exposes the result's outer constructor or
function closure, not necessarily its contents.

The target has four states: descending into a source subtree, applying an
algebra to a completed layer, returning a child result to a frame, and
finishing at the root. At descent, it enters the first recursive field and
saves a frame; when no field remains, it exposes an algebra request. On a
child return, it either enters the next child with an updated frame or exposes
the completed layer. Each transition is a nonrecursive, constructor-specialised
clause. A shared tail-recursive driver invokes those clauses and forces each
successor state; saved source fields remain lazy.

\subsection{Suspension and operation contracts}
The generated interface provides \code{start}, \code{step}, \code{pending},
\code{runFor}, and \code{run}. \code{start} saves the source without inspecting
it; \code{step} executes one transition, leaving a finished state unchanged,
and \code{run} drives the machine to completion. The operation \code{runFor}
executes at most its positive fuel budget and returns either
a completed result or a resumable state with the fuel consumed.
\code{pending} exposes an algebra request without invoking the callback. A
suspended state may be retained and resumed more than once with the same pure
algebra.

The contracts distinguish descent, which demands the current outer
constructor; application, which forces only the algebra result to weak head
normal form; and return, which demands neither source nor algebra. Each has
bounded structural overhead for a fixed description. Source-thunk evaluation
and client-algebra work are charged separately, so one transition cannot
interrupt a diverging thunk or callback. The JSON records this separation and
the frame layouts; it is explanatory metadata, not an independently checked
proof object.

\begin{theorem}[Indexed family fold and transition budget]\label{thm:fold}
For a finite value of any accepted family root with $N$ constructor
occurrences across all family sorts, suppose every
constructor demand and algebra evaluation required by the reference eager
fold terminates. The emitted machine returns the same result and has the
same ordered constructor demands and algebra applications. From
\code{start} to \code{Finished}, it executes exactly $3N$ transitions:
$N$ descents, $N$ applications, and $N$ returns. Its structural overhead is
$O_{\boldsymbol\Phi}(N)$, excluding source forcing and algebra work. Partitioning the
execution into positive fuel budgets preserves its result and total
transition count. Each live history contains one frame per suspended
ancestor.
\end{theorem}
\begin{proof}
Represent each pending reference call by its parent sort, constructor,
active child sort and field, completed results, and unvisited fields. The
single-sort Equation~\eqref{eq:fold-frame}, or its indexed family form
above, gives exactly this record. Relate the reference call stack to its
backward trail in the same ancestor order; Theorem~\ref{thm:family-trail}
ensures that the sort indices compose. Descent saves the record for the first
child. On a child return, the emitted clause either replaces that record
with the one for the next child, or constructs the parent layer from all
completed results. Application invokes the same algebra on that layer and
forces its result to the same extent. A return with empty history completes
the root. These cases preserve the stack relation and the order of source
demands and algebra calls. Induction on the finite source proves agreement and termination.

Each constructor is entered once, its completed layer is applied once, and
its result is returned once, including the root return. This gives $3N$.
A transition inspects and constructs only bounded-size records for fixed
$\Phi$. The driver executes at most the supplied positive fuel and then
returns its current state without modifying it, so concatenating resumed
chunks gives the same transition sequence. The stack invariant also gives
the history bound. Retained subtrees and result payloads are not bounded
by the number of control records.
\end{proof}

For potentially infinite data, every finite execution prefix still has a
finite history. A finite positive budget returns a suspension or a completed result whenever
each demand and application it reaches terminates. This does not assert
that the fold will ever finish or produce a useful result on an infinite
input. For example, an infinite list admits arbitrarily long finite
descent prefixes, but this eager fold never reaches its empty constructor.
Corollary~\ref{thm:productivity} gives finite observations for cursors; this eager fold has no such guarantee.

A lazy binary fold with branch algebra $\lambda x\,y.\,x$ can return
from \code{Branch (Leaf 42) undefined}. Our eager fold enters the right
child and fails to return. Preserving such short-circuiting, including
sharing and the algebra's demand for recursive results, requires another
translation and correctness argument. Likewise, the explicit machine
moves traversal continuations from the host call stack into heap records;
it does not promise constant total memory, bounded callback stack use, or
backwards navigation through results that the algebra has discarded.

\begin{theorem}[End-to-end compilation and contracts]
\label{thm:end-to-end}
For every accepted ordered family $\boldsymbol\Phi$,
the compiler $\mathsf{compile}$ applied to $\boldsymbol\Phi$ terminates and emits cursor and eager-fold code, whose types are the indexed schemas of Section~\ref{sec:arden} and Appendix~\ref{app:folds}, together with a contract report. The ordinary cursor
operations satisfy their reported contracts in the relation of Appendix~\ref{app:contract-semantics}. For every finite total root on which the stated eager
evaluation terminates, the completed fold run satisfies the reported
$3N$ transition count, constructor-demand trace, and client-algebra trace.
Any emitted cursor certificate has the finite-observation guarantee of
Theorem~\ref{thm:family-prefix}, or the local special case in
Appendix~\ref{app:local-guard}.
\end{theorem}
\begin{proof}
The source description is finite, and each compilation pass enumerates only
its constructors, field positions, and finite lowering closure. Theorem~\ref{thm:correctness}, Corollary~\ref{thm:productivity}, and Theorems~\ref{thm:entry-contract}, \ref{thm:move-contract}, \ref{thm:cursor-contract}, \ref{thm:cost}, \ref{thm:family-prefix}, and~\ref{thm:guard} establish the cursor clauses and their reports.
Theorem~\ref{thm:fold} establishes the indexed eager-fold code, its
transition count, and its traces. The generated types are the
indexed derivative, trail, and refined-frame schemas used in those results;
GHC acceptance of the emitted modules is tested, not proved.
\end{proof}

\subsection{Eager-fold validation}
\label{sec:fold-validation}
The fold suite of Appendix~\ref{app:counts} also tests boundary suspension, undefined fields, a
10,000-transition prefix, and a four-recursive-field datatype.

With GHC optimisation disabled (\code{-O0}), a depth-200,000 skew binary tree
completes under a 1\,MiB stack
limit with unbounded and 257-transition drivers. It has 400,001 constructors,
executes 1,200,003 transitions, and sums 200,001 unit leaves. This tests
control-stack safety for this scalar algebra, not arbitrary client code or
allocation. Appendix~\ref{app:fold-runtime} gives supplementary runtime
comparisons against handwritten and recursive folds.

\subsection{Supplementary eager-fold measurements}
\label{app:fold-runtime}
The fold experiment compares four implementations of annotation summation on
a fully constructed depth-4,096 skewed syntax tree (8,193 constructors). The
generated fold runs unbounded or in 31-transition budgets; baselines are a
handwritten stack and direct recursion. A difference-list check instantiates
Lemma~\ref{lem:batch-semantics} for \code{ordered-elements}.
\begin{table}[htbp]
\centering
\caption{Finite eager-fold batch: annotation sum over a depth-4,096 skew
syntax tree. Time and allocation are medians per tree.}
\label{tab:fold-runtime}
\begin{tabular}{lrr}
\hline
Policy & \(\mu\)s/tree & KiB/tree\\
\hline
\kamot{} eager fold & 216 & 671\\
\kamot{} fold, fuel 31 & 238 & 894\\
Handwritten explicit stack & 209 & 512\\
Direct recursive fold & 173 & 160\\
\hline
\end{tabular}
\end{table}

Table~\ref{tab:fold-runtime} gives the medians. Direct recursion is fastest on this finite, available input. The unbounded
generated fold, a top-level loop over \code{step} that GHC specialises to the
datatype, takes 216\,$\mu$s, within 4\% of the handwritten stack's
209\,$\mu$s, but allocates 671\,KiB rather than 512\,KiB. Running it in
31-transition budgets costs 238\,$\mu$s and 894\,KiB. The generated machine
provides suspension at higher allocation than either baseline.

\section{Demand-admission measurements}
\label{app:admission}
The admission study uses the syntax family from
Appendix~\ref{app:family-service}. Each input contains a \code{Let}, a
binding, a chain of \code{App} nodes, and a final expression. The local
workload samples eight positions in chains of depth zero to three; the
deep-local workload samples eight positions in a depth-96 chain; and the
long-return workload starts at the deepest leaf in chains of depth 24--128.
The next label in the last case lies beyond pending argument lists, whose
constructors must be demanded on return.

A pure, non-inlined producer performs 128 fixed \code{Word64} mixing
iterations before exposing each constructor; its children remain lazy.
Preparing the initial cursor and its old prefix is outside timing. An
instrumented build checks that preparation demands exactly the prefix
predicted by the independent source interpreter, and that subsequent entry
events correspond to fresh producer calls. The timed executable has no
producer instrumentation. This models a source with nontrivial production
cost, not a measured parser or I/O system.

\emph{Fuelled} checks transition and demand credits before each step.
\emph{Certified} starts a move only if its emitted bounds fit both credits,
using the terminal suffix's bound when the move has no return outcome and the depth bound
otherwise. \emph{Defer} skips the first window; \emph{direct} ignores both
credits. All survivors use the same fuelled driver. Three workloads cover
shallow local queries, cheap deep queries, and long returns through child
lists. Across budgets and cancellation rates, we measure 108 configurations
checking answers, budgets, and demands against
independent traces.

\begin{table}[htbp]
\centering
\caption{Admission on 256 long-return queries; 50\% of unfinished queries
are cancelled after the first window. Work and Demand are per-window
transition and constructor credits;
Wasted counts constructor evaluations for cancelled queries. Time and
allocation are medians per submitted query, including cancelled queries.}
\label{tab:admission}
\begin{tabular}{rrlrrrr}
\hline
Work & Demand & Policy & Answers & Wasted & $\mu$s & bytes\\
\hline
64 & 8 & Fuelled & 128 & 1,024 & 9.54 & 18,151\\
64 & 8 & Certified & 128 & 0 & 8.77 & 16,841\\
256 & 32 & Fuelled & 143 & 3,616 & 12.08 & 22,018\\
256 & 32 & Certified & 128 & 0 & 8.55 & 16,007\\
512 & 256 & Fuelled & 256 & 0 & 16.52 & 30,077\\
512 & 256 & Certified & 243 & 0 & 14.48 & 27,605\\
\hline
\end{tabular}
\end{table}

The application generates an admission helper from the emitted JSON. The
terminal suffixes of \code{App}, \code{Bind}, and \code{Let} can only yield;
their work/demand bounds are $(2,1)$, $(2,1)$, and $(4,2)$ respectively.
The suffix of \code{Lit} returns, so the helper uses the family bound
$(5+3h,2+2h)$. It inspects only available terminal tags and cached depth,
and rejects arithmetic overflow. Neither measured costs nor oracle traces
are inputs to admission. An initial run used only the family bound; both runs and their exact generated sources are retained.

We cross three budget pairs, $(64,8)$, $(256,32)$, and $(512,256)$, with
cancellation rates 0\%, 50\%, and 100\%, three workloads, and four policies.
Demand credits are the tighter admission constraint in this study.
The rate selects query identifiers by
$(37\mathit{id}+11)\bmod100<\mathit{rate}$, independently of policy. An
edit callback after the first window cancels only a query still unfinished; answers
already delivered remain useful. The direct control may delay that callback
by overrunning its quotas. A deferred survivor retries with the same unboxed
driver as a fuelled survivor. Callback bookkeeping is measured, but the
callback performs no other foreground computation.

The application is built with GHC~9.12.1 at \code{-O2}, without \code{-threaded}, and runs with \code{+RTS -T -A32m} on the machine of Section~\ref{sec:applications}. Seven fresh processes measure each configuration, rotating the order and
retaining every sample. A throughput process runs 256 queries without
per-window clocks or event logs; separate processes time 64 queries window
by window. The artifact records 756 throughput samples, 78,533 windows, and
648 summary rows. Its validator recomputes all counters and window traces
from the source interpreter, checks source hashes and regeneration, and
reproduces every summary. An instrumented check covers all 108 configurations,
zero-fuel non-strictness, and a bottom source that remains unforced after
demand credits run out.

For the small-budget, 50\%-cancellation long-return case, fuelled service
takes 9.20--9.73\,$\mu$s per submitted query and certified admission takes
8.38--9.11\,$\mu$s. Both deliver the same 128 answers. Their total demands
are 10,760 and 9,736: the difference is exactly the 1,024 discarded producer
calls. Direct search delivers all 256 answers but exceeds the quota in every
first window. Deferring everything matches certified admission on this
workload but loses half of the local answers. Fuelled and certified admission
answer all 256 local queries. At 32 demand credits, the terminal-suffix bound
recovers 93 deep-local answers lost when admission uses only depth. On long
returns with budgets $(256,32)$, certified admission saves 3,616 discarded
evaluations but loses 15 answers that fuelled search completes; at
$(512,256)$ it loses 13.

The no-cancellation control exposes the cost of waiting. On the same long-return inputs at budget $(64,8)$, every policy produces all 256 answers and demands all
19,200 constructors. Certified admission adds 256 callbacks and allocates
32,571 bytes per query, against 31,925 for fuelled execution; their timing
ranges overlap. With 50\% cancellation at larger budgets, Table~\ref{tab:admission} reports useful answers lost through conservative deferral. The 1\,$\mu$s clock is too
coarse to resolve sub-microsecond local windows. Allocation measures total
service allocation, not retained space, and no structural bound here is a
wall-clock guarantee.

\section{Scheduling measurements}
\label{app:scheduling}
The scheduling study compares three policies with budget $\beta=128$
transitions on preconstructed rose trees. Direct search runs each move to its
yield; fuelled search always runs a fuelled slice; contract-directed search
runs the move directly when its certified bound $b_d(h)$ is at most $\beta$
and runs a fuelled slice otherwise (Corollary~\ref{cor:scheduling}). Its three inputs expose different costs.
The balanced four-way tree of depth six has 5,461 labels and short moves. The
deep-return tree places a chain of 8,193 labelled nodes before a sibling,
producing a long return through saved frames. The sparse variant adds a
label-free constructor; 64 labelled subtrees each sit behind 256 silent nodes,
and its prefix cycle receives no certificate.

Throughput samples scan 64 roots with distinct label seeds; separate fresh
processes time individual scheduler slices. There are seven samples per
workload and policy, with rotating policy order and the GHC/RTS configuration
of Section~\ref{sec:applications}. Checks compare outputs and complete
transition counts before timing and verify every certified bound and fuel
limit. The experiment measures a cooperative scanning kernel, not end-to-end
editor response time. Table~\ref{tab:scheduling} reports the results.

\begin{table}[ht]
\centering
\caption{Family-cursor scheduling. Time and allocation are medians per label.
The final column is the median of seven per-process maximum slice durations,
in $\mu$s. Maximum step counts are deterministic.}
\label{tab:scheduling}
\begin{tabular}{llrrrr}
\hline
Input & Policy & ns/label & bytes/label & Max steps & Max slice\\
\hline
Balanced & Direct & 66.8 & 895 & 46 & 4\\
 & Fuelled & 68.5 & 919 & 46 & 2\\
 & Contract-directed & 66.8 & 895 & 46 & 3\\
Deep return & Direct & 123.8 & 895 & 32,774 & 356\\
 & Fuelled & 130.8 & 920 & 128 & 5\\
 & Contract-directed & 127.5 & 920 & 128 & 3\\
Sparse & Direct & 17,030 & 178,189 & 1,542 & 20\\
 & Fuelled & 17,663 & 178,497 & 128 & 4\\
 & Contract-directed & 18,657 & 178,497 & 128 & 6\\
\hline
\end{tabular}
\end{table}

On the deep-return input, the seven per-process maxima range from 325 to 857\,$\mu$s for direct search and from 3 to 19\,$\mu$s for contract-directed
search. The throughput ranges of the direct, fuelled, and contract-directed
policies overlap: 111.8--134.9, 115.3--155.4, and 118.4--146.9\,ns per label. On the balanced input the
ranges are 63.3--69.8, 67.6--74.6, and 64.7--69.9\,ns. The depth bound is
conservative, so many individually cheap moves on the deep tree also take the
fuelled path. A direct move avoids the fuelled driver's boxed slice result,
which saves 24 bytes per slice, or 2.6\% of allocation on the balanced input. The admission study uses
an unboxed driver, so its avoided source production does not depend on that
overhead. The observed clock quantum is 1\,$\mu$s, so small differences between shallow
slice maxima are not meaningful. Allocation is measured, but collection
counts and pauses are not; maximum slice times may include collections.

\section{Additional measurements}
\label{app:cursor-evaluation}
\subsection{Clause counts and validation coverage}
\label{app:counts}
For the precision study, an independent checker validates each productive
divergence witness. Fields that the divergent search should never visit hold
sentinel bottoms, so a search that takes a wrong path fails. Replacing the
sentinels with finite values gives fully productive inputs on which the first
20,000 transitions are unchanged. Divergence follows from the prefix-cycle
argument; the finite traces test the implementation of each witness.

Nine descriptions generate ordinary and instrumented modules, checked with
GHC~9.12.1 using warnings as errors. The cursor and cost suites cover 4,978
finite values (including 1,188 with mixed fields) and 28,482 positions.
The fold suite checks 5,063 finite inputs against independent recursive
specifications, the $3N$ count, all transition kinds, and budgets 1, 2, 7,
and 31; Appendix~\ref{sec:fold-validation} gives its further tests.

Table~\ref{tab:counts} lists measured clause counts for complete traversals
of four inputs; each satisfies $q=2N+M$ (Equation~\eqref{eq:cost}). Its
Largest search column ranges over initial entry and all moves, including
the final boundary request. The sparse
tree is \code{Split (Item 0) (Split e (Item 1))} over the \code{Sparse} type of
Section~\ref{sec:cost}, where \code{e} is a complete tree of \code{Empty} and
\code{Split} nodes of height 10. One move takes 4,098 clauses; the value fails
the local entry condition (Appendix~\ref{app:local-guard}).
\begin{table}[ht]
\centering
\caption{Measured abstract clause counts for complete forward traversals.
Every row satisfies $q=2N+M$.
These are deterministic structural counts, not time measurements.}
\label{tab:counts}
\begin{tabular}{lrrrr}
\hline
Input & $N$ & $M$ & Total $q$ & Largest search\\
\hline
Log, 1,000 events & 1,001 & 1,000 & 3,002 & 1,003\\
Balanced syntax, height 9 & 1,023 & 1,023 & 3,069 & 11\\
Skewed syntax, 512 applications & 1,025 & 1,025 & 3,075 & 4\\
Sparse tree, empty height 10 & 2,051 & 2 & 4,104 & 4,098\\
\hline
\end{tabular}
\end{table}

\subsection{Replay and correction of a lazy event log}
An event log may have a finite observed past and an unbounded future:
\begin{lstlisting}
data Log a = End | Event Integer a (Log a)
\end{lstlisting}
The integer is an opaque timestamp and \code{a} is the editable payload.
Differentiation gives a frame containing the timestamp and old payload,
and a terminal record containing the timestamp and future log. The
constructor is forward entry-guarded. Opening a nonempty log yields its
payload without forcing the tail; each successful forward move inspects
only the next event constructor. Successful adjacent moves have constant
structural cost for this datatype. The history permits replay
backwards and correction of one observed payload.

The streaming repair workload opens the first event, makes 2,048 forward
requests, corrects every 64th payload, and retains the cursor for backward
replay. A separate application test replaces event 999 and closes the
edited cursor. The checks establish persistence, replay, and prefix repair;
opening \code{Event 0 0 undefined} leaves the suffix unforced.
The generated machine executes 4,097 clauses: one entry and two per successful
forward move. We compare it with a productive traversal-based generic zipper
and root rescan; timing covers advancing, repairing, and closing the prefix.

Table~\ref{tab:stream-runtime} gives seven-sample medians in the configuration
of Section~\ref{sec:prototype}. The generated cursor is six times faster than the generic zipper
and allocates under one quarter as much per request; rescan has no replayable
history and is about 114 times slower. The dissection port is evaluated on the
finite annotation workload, because it has no packaged coinductive frontend.
\begin{table}[htbp]
\centering
\caption{Infinite streaming repair: 2,048 forward requests and a repair every
64 requests. Time and allocation are per-request medians; separate live-heap
runs observe the same 2,050-event prefix.}
\label{tab:stream-runtime}
\begin{tabular}{lrrr}
\hline
Policy & ns/request & bytes/request & live KiB\\
\hline
\kamot{} cursor & 25 & 247 & 202\\
External \code{zippers} & 152 & 1,216 & 919\\
Direct recursive rescan & 2,839 & 1,332 & 170\\
\hline
\end{tabular}
\end{table}

\subsection{Incremental annotation service}
The annotation-service kernel edits annotations attached to syntax nodes:
\begin{lstlisting}
data Expr a
  = Var a Int
  | Lit a Integer
  | App a (Expr a) (Expr a)
  | Lam a Char (Expr a)
\end{lstlisting}
The focus is an annotation, not an arbitrary subtree. Every recursive
constructor begins with an annotation, so forward traversal satisfies the
local entry condition of Theorem~\ref{thm:guard} with $f=2$. A move from depth $h$ consequently uses
at most $5(h+1)+1$ search clauses. This applies even to an incrementally
unfolded infinite syntax value, provided its demanded constructors are
productive. A client can visit and update available annotations without
forcing a variable identifier, literal, or unvisited sibling.

Tests cover balanced and skewed trees, finite reconstruction, undefined
unvisited fields, and a thousand moves through an infinite annotated unfolding.
The incremental annotation-service kernel answers 2,048 successive ``next annotation''
requests and persists an edit every 64 requests. Its standalone selector is the one of Appendix~\ref{app:selection}; for this
profile, productive local replacement, it selects the persistent cursor.

On a skewed syntax value of depth 4,096, the selected cursor executes 4,097
search clauses; restart executes 4,196,352. Both return the same annotations,
and the cursor reconstructs the edited root. Thus the generated contract
justifies retaining the typed continuation for sequential interaction with
local edits.
\subsection{Runtime and allocation comparison}
\label{app:runtime-comparison}
We measured this service on a fully constructed depth-4,096 skewed syntax tree.
Table~\ref{tab:runtime} reports seven-sample medians in the \code{-threaded}
configuration of Section~\ref{sec:applications}. Allocation is the RTS \code{allocated\_bytes} delta;
each timing sample runs in a fresh process.
The experiments use no core pinning or frequency control. Samples run
16 roots for recursive rescan, 64 for each generic baseline, and 256 for
the generated cursor and handwritten stack. Reported ranges describe
fresh-process variation, not confidence intervals.
Times are whole-session measurements divided by request count, not individual
request latencies. The dissection harness also converts the completed
polynomial representation back to the native tree inside the measured interval;
its initial conversion is outside that interval. Timings include navigation,
reconstruction, and output conversion.

A separate run of seven fresh-process samples on 64 preconstructed roots stops the
dissection session after navigation and edits, after reconstruction and
observation of the polynomial result, or after conversion to the native tree.
The table reports per-request medians and timing ranges.
\begin{center}
\begin{tabular}{lrrr}
\hline
Stopping point & ns/request & Range & bytes/request\\
\hline
Navigation and edits & 270 & 252--329 & 482\\
Polynomial result & 613 & 591--628 & 850\\
Native tree & 604 & 555--669 & 1,058\\
\hline
\end{tabular}
\end{center}
Conversion adds 208 bytes per request. Its time cost cannot be isolated: the
difference between the two longer modes lies within run-to-run variation, and
the two modes observe different trees. Navigation and edits alone take 2.5
times the generated cursor's complete session time.

Heap snapshots use separate fresh processes. For finite inputs, every policy
fully observes the source and reconstructed output before collection. A
snapshot retains one cursor or edited root; it reports total live heap,
including reachable source data. Stream snapshots materialise and observe
exactly 2,050 events; stream timing starts with only the first event ready.
All stream policies use the same source datatype.

Exact-output checks precede measurement, and a test with an undefined stream
tail detects demand beyond the observed prefix. The artifact records all
samples, their ranges, source hashes, and build and runtime settings.

The generated cursor has similar amortised time per request to the handwritten
stack (107 versus 113\,ns) while allocating 24\% fewer bytes. In this harness,
it is 5.2$\times$ faster than the generic-dissection port and uses about 19\%
of its allocation; against \emph{Jacket}/\emph{Path}, it is 6.4$\times$ faster
and uses about 5\% of its allocation. Root rescan is 46$\times$ slower and allocates 95$\times$
more per request.
Retaining the generated cursor costs about 32\,KiB more live heap than retaining
the rescan root in each workload.

\subsection{Syntax-service family kernel}
\label{app:family-service}
The family
\begin{lstlisting}
data Expr a = Lit a | App a (Expr a) [Expr a] | Let a [Decl a] (Expr a)
data Decl a = Bind a (Expr a)
\end{lstlisting}
lowers its argument and binding lists to auxiliary sorts. Forward moves are
certified with at most $5+3h$ transitions and $2+2h$ constructor demands;
backward moves are unknown. The session makes 2,048 next requests from the
first annotation, replaces every 64th selected annotation, then closes and
observes the rebuilt root. The mixed input interleaves applications and lets
and has 23,206 annotations; the deep input is a chain of 2,048 lets with
6,145 annotations. Two handwritten machines use the same lowered source
types. The indexed machine has independently written GADT frames, navigation,
and reconstruction, with non-strict payloads matching the generated cursor.
The unindexed machine has a sum-of-sorts representation,
dynamic sort checks, and strict payload fields. Rescan searches from the root
for every request. Every policy's output matches an independent preorder
oracle before timing. Each input and policy has seven fresh-process samples
with rotating order, over 256 roots per process for all four policies.

\begin{table}[htbp]
\centering
\caption{Syntax-service family kernel: medians per request.}
\label{tab:family-service}
\begin{tabular}{lrrrr}
\hline
& \multicolumn{2}{c}{Mixed} & \multicolumn{2}{c}{Deep let chain}\\
Policy & ns & bytes & ns & bytes\\
\hline
\kamot{} family cursor & 78 & 246 & 58 & 253\\
Handwritten indexed stack & 78 & 246 & 56 & 253\\
Handwritten unindexed stack & 87 & 357 & 38 & 309\\
Rescan & 3,757 & 16,384 & 3,396 & 11,308\\
\hline
\end{tabular}
\end{table}

Generated and handwritten indexed cursors have identical per-request
allocation (Table~\ref{tab:family-service}). Their timing
ranges overlap on both inputs: 77--83 versus 78--85\,ns on the mixed tree,
and 56--59 versus 55--58\,ns on the let chain. The unindexed machine is faster
on the let chain but allocates more; these differences depend on its
representation. The artifact also retains an earlier three-policy run, whose samples are not pooled with these. Additional checks cover
36 small roots at every request count, final boundaries, partial payloads,
and an infinite prefix; these are correctness checks, not timing inputs.

\subsection{Code size and compile time}
\label{app:family-scaling}
Across 12 one-sort descriptions varying constructor count and recursive-field arity,
generated Haskell ranges from 1.9\,KiB (87\,ms to check) to 30.3\,KiB
(159\,ms).

The scaling run covers the 19 corpus datatypes after container lowering
(1--10 sorts, 2--21 constructors) and the 81 generated families of
Section~\ref{sec:precision} (1--3 sorts, 3--15 constructors). Cursor modules
range from 2.9 to 29.0\,KiB, resumable cursor modules from 6.2 to 45.6\,KiB,
and fold modules from 3.0 to 18.8\,KiB. A least-squares fit of cursor size
against active fields gives about 1.8\,KiB plus 0.6\,KiB per active field
($R^2=0.96$); constructor count alone fits less well ($R^2=0.81$). GHC \code{-fno-code} takes 87--149\,ms per cursor module
and 98--175\,ms per fold module, including the source module it imports. The
largest module comes from the located Haskell-module datatype (10 sorts, 21
constructors).

These corpus fits do not establish an asymptotic scaling law. A second study
uses rings of 1, 4, 16, and 32 mutually recursive sorts, with a labelled leaf
and a labelled binary constructor per sort. Two further cases independently
increase constructor count or recursive-field arity. Every measurement is a
fresh GHC invocation with a fresh output directory and
\code{-fforce-recomp}; we report medians of three samples. Cursor modules
range from 3.8 to 221.6\,KiB and take 0.16--2.76\,s to compile with
\code{-O2}. Fold builds, including their generated source module and runtime
dependency, take 0.27--3.46\,s. The rise in cursor size reflects reconstruction
helpers repeated for each root sort, as accounted for in
Section~\ref{sec:transitions}. The 32-sort fold exceeds GHC's default
pattern-checking model budget of 30; the complete study uses
\code{-fmax-pmcheck-models=256} throughout, with \code{-Wall -Werror}.
The initial failed run is retained in the artifact. This is a controlled
build-cost study, not evidence about whole-program optimisation or runtime.

\section{Proof of compositional family contracts}
\label{app:proof-family-prefix}
\begin{lemma}[Empty returns witness nullability]
\label{lem:empty-return}
If a reference run enters a value $t$ of sort $i$ and later returns from $t$
without an intervening yield, then $\mathsf{Null}(i)$.
\end{lemma}
\begin{proof}
The run between the entry and the return is finite. It leaves each node it
enters only after scanning all of that node's active fields, and it does not
yield, so no entered node has an $X$ field and every $Y$ field of an entered
node is itself entered and returned from. The entered nodes form a finite
tree rooted at $t$. Induction on its height, applying the
nullability rule at each node, shows that every entered sort, in particular
$i$, is nullable. Constant fields are never entered, so they may hold any
value.
\end{proof}

\begin{lemma}[Suffix summary soundness]
\label{lem:suffix-summary}
Let $u$ be any ordered suffix of a constructor's fields with a bounded summary
$S(u)$, and assume demanded roots are productive. Suppose every entry summary
consulted in computing $S(u)$ is sound under this assumption:
entering any value of its sort reaches a yield or a pending return within the
corresponding outcome bound, and cannot take an omitted outcome. Starting the
scan at $u$, with any well-sorted field values and stack, reaches a yield or
the surrounding constructor's pending return within the corresponding bound
of $S(u)$. The bound excludes the control action that starts the scan. No condition
is imposed on the fields preceding $u$.
\end{lemma}
\begin{proof}
Induct on the length of $u$. The empty suffix returns and an element field
yields, each with cost $(0,0)$; skipping a constant changes neither cost nor
outcome. For $u=Y_j u'$, soundness of $\mathcal E_j$ gives two possibilities.
A child yield ends the scan within $\mathcal E_j.\mathsf{yield}$. A child
return costs at most $\mathcal E_j.\mathsf{return}$ before its pending return
action; that action costs $(1,0)$ and resumes at $u'$, to which the induction
hypothesis applies. These are exactly the alternatives in the rules for
$S(Y_j u')$. An omitted child outcome cannot occur, so an alternative that
uses it need not be bounded. If an alternative uses no omitted child outcome, every summary needed for $u'$ was consulted in computing $S(u)$, so the induction hypothesis applies. Componentwise maximum bounds either yield
alternative, and addition bounds the return alternative. Thus omitted suffix
outcomes cannot occur either.
\end{proof}

\begin{proof}[Proof of Theorem~\ref{thm:family-prefix}]
Write $\sqsubset$ for the order in which bounded entry summaries are
evaluated, so $\mathcal E_i$ depends only on summaries $\mathcal E_j$ with $i\to_d j$ and
$j\sqsubset i$. A sort without constructors has no values, and both of its
outcomes are impossible.

\emph{Claim A.} Let $\mathcal E_i$ be bounded, $t$ a finite or infinite value of sort $i$, and $\kappa$
any well-sorted stack. The run from $\mathsf{Enter}_d(t,\kappa)$ reaches either a
$\mathsf{Yield}$ within $\mathcal E_i.\mathsf{yield}$ or the configuration
$\mathsf{Return}_d(t,\kappa)$ within $\mathcal E_i.\mathsf{return}$, counting
transitions and constructor demands, and only an outcome that $\mathcal E_i$ marks
possible occurs.

We prove Claim A by induction on $\sqsubset$. Let $t=c(\mathbf v)$. The entry
step matches $c$ and costs $(1,1)$. For the scan of $c$'s fields we prove
\emph{Claim B}: for each suffix $u$ of the fields of $c$ in direction $d$ that
the scan can reach without yielding, it reaches a $\mathsf{Yield}$ within
$S(u).\mathsf{yield}$ or the end of $c$ within $S(u).\mathsf{return}$.

We induct on the length of $u$. The empty, element, and constant cases are as in the proof of Lemma~\ref{lem:suffix-summary}. Suppose $u=Y_j\,u'$.
The scan reached this field without yielding, so every earlier active field of
$c$ is a $Y$ field whose child returned empty, and by
Lemma~\ref{lem:empty-return} each of those sorts is nullable. The edge rule
therefore gives $i\to_d j$, so $j\sqsubset i$ and Claim A holds for the child
$t'$ of sort $j$.

The child segment either yields within $\mathcal E_j.\mathsf{yield}$, which ends
the scan, or reaches its pending return within $\mathcal E_j.\mathsf{return}$. In the
latter case the return step costs $(1,0)$, demands nothing, and resumes the
scan at $u'$, where the induction hypothesis applies. This is the rule of Section~\ref{sec:prefix-composition} for $S(Y_j\,u')$. An outcome marked impossible cannot occur, because each
of its alternatives contains an impossible child outcome or an impossible
suffix outcome. Adding the entry cost and maximising over constructors gives
$\mathcal E_i$. The child $t'$ may be infinite or contain elements: then it cannot return empty, by the proof of Lemma~\ref{lem:empty-return}, and Claim A bounds its
segment up to its first yield.

\emph{Moves.} By Lemma~\ref{lem:mixed-operations}, a valid cursor at depth $h$ encodes a yield whose stack holds $h$ records. A move from a cursor whose terminal context records constructor
$c$ and position $\ell$ performs one advance step and scans the fields after
$\ell$. This suffix need not be reachable from constructor entry before a
yield, so Claim B alone does not apply. The emitted move certificate requires
bounded summaries for every terminal and frame suffix. Claim A has established
soundness of every entry summary they consult, so
Lemma~\ref{lem:suffix-summary} applies at each of these starting positions.
Adding the advance action gives the terminal bound $(w_B,n_B)$. If this
segment reaches the pending return from $c$, returning through the nearest
frame starts its suffix; including that return action gives $(w_K,n_K)$.
Each frame segment starts by popping one frame of the original stack. Any child
entered within that segment either yields or returns before the segment
ends, so an exhausted segment leaves no new frames. Hence at most $h$ such frame segments run, and the
final root return to $\mathsf{End}$ costs one further step with no demand.
Summing gives $b_d(h)=w_B+h\,w_K+1$ and $n_B+h\,n_K$. Entry is Claim A at the
root with the same final step.

Each step demands at most one outer constructor, which a productive root
supplies in finite time, so a bounded number of steps takes finite time.
Lemma~\ref{thm:simulation} maps each reference step to one generated clause
with the same demand, which transfers the bounds to the generated searches.
\end{proof}

\paragraph{Rose-tree refinement.}\label{app:rose-depth}
For a focus in a value of sort \code{Tree}, each of its $\delta$ ancestors
contributes one \code{Frame\_Tree\_Node\_1} and one
\code{Frame\_List0\_List0Cons\_0}. Each earlier sibling contributes one
\code{Frame\_List0\_List0Cons\_1}, so $h=2\delta+s$ for $s$ such siblings.
Their return outcomes are $(1,0)$, $(3,1)$, and $(1,0)$, respectively; the
terminal return outcome is $(3,1)$. If every segment returns, the move costs
at most $(4+4\delta+s,1+\delta)$, including the final root return.
A yielding segment replaces its return outcome $(3,1)$ by $(3,2)$ and
skips the remaining frames and root return. Thus every move is bounded by
$(4+4\delta+s,2+\delta)$: only the demand bound depends solely on source
depth; the work bound also counts earlier siblings.

\section{Proof of completeness}\label{app:proof-completeness}
Throughout, the family is inhabited and accepted, and $d$ is fixed. By
Lemma~\ref{thm:simulation} it suffices to count reference steps.

\begin{lemma}[Element-free values]\label{lem:element-free}
For every nullable sort $j$ there is a finite value $z_j$ of sort $j$ that
contains no element, and a run entering $z_j$ returns from it after finitely
many steps without yielding.
\end{lemma}
\begin{proof}
Induct on the derivation of $\mathsf{Null}(j)$. Its last rule names a
constructor with no $X$ field whose recursive fields have nullable sorts; fill
those with the values given by induction and each constant field with a value
of its type. The run scans every field of every node it enters, finds no
element, and terminates because the value is finite.
\end{proof}

\begin{lemma}[Realising prefix edges]\label{lem:edge}
If $i\to_d j$, then for every value $t'$ of sort $j$ there is a value $t$ of
sort $i$, finite apart from $t'$, whose entry reaches $t'$
after at least one step and without yielding.
\end{lemma}
\begin{proof}
The edge rule names a constructor $c$ of sort $i$ and a field $\ell$ with
$E_{c,\ell}=Y_j$ such that every active field before $\ell$ in direction $d$
is a $Y$ field of nullable sort. Let $t=c(\mathbf v)$ with $v_\ell=t'$, the
element-free values of Lemma~\ref{lem:element-free} in the earlier active
fields, any values in constant fields, and finite values in the later fields;
these exist because every sort has a finite value. The run matches $c$,
skips constants, enters and returns from each earlier child without
yielding, and then enters $t'$.
\end{proof}

\begin{lemma}[Pumping]\label{lem:pumping}
If $\mathcal E_j$ is unknown, then for every $n$ there is a finite value
$t_n$ of sort $j$ such that a run entering $t_n$ takes at least $n$ steps
without yielding, and there is a productive infinite value $t_\infty$ of sort
$j$, whose constructors are all defined, on which a run entering it neither
yields nor returns.
\end{lemma}
\begin{proof}
Entry summaries are evaluated while some sort has all its prefix successors
evaluated. A sort left unknown therefore has a prefix successor that is also
unknown. Since there are finitely many sorts, we can choose an infinite walk
$j=j_0\to_d j_1\to_d\cdots$ consisting of a finite stem followed by a
repeated cycle. Applying Lemma~\ref{lem:edge} along its first $n$ edges,
with any finite value at the end, gives $t_n$; each edge costs at least one
step and no yield. Construct $t_\infty$ along the whole walk, including the
stem, using the edge constructors and finite fillers from that lemma.
Every constructor is defined. The run follows the infinite walk, so it
neither yields nor returns from $t_\infty$.
\end{proof}

\begin{lemma}[Reachability of finite contexts]\label{lem:context-reachability}
Every well-sorted cursor whose reconstructed root has a finite, fully
defined constructor tree is reached by entry and finitely many moves in
direction $d$ on that root.
\end{lemma}
\begin{proof}
Fill the terminal hole with the focused element, then fill each frame's
recursive hole with the reconstructed child, nearest frame first. Each
filling preserves its constructor's field order and sort, yielding a finite
root. Traversal of a finite child visits its elements and returns. Thus, at
each ancestor of the focus, finitely many steps and advances traverse the
fields preceding the chosen child in direction $d$ and enter it with exactly
that ancestor's frame saved. Induction down the path reaches the terminal
constructor with the given trail. Traversing its preceding fields in
direction $d$ then reaches the focus with the given terminal record.
Grouping the steps between successive yields gives entry followed by
completed public moves.
\end{proof}

\begin{proof}[Proof of Theorem~\ref{thm:completeness}]
The forward implications of parts~1 and~2 follow from
Theorem~\ref{thm:family-prefix}.

\emph{Part 1.} If $\mathcal E_r$ is unknown for a root sort $r$, root entry on
the value $t_n$ of Lemma~\ref{lem:pumping} takes at least $n$ steps.

\emph{Part 2.} Suppose no move certificate is emitted. Some terminal or frame
suffix $u$, following field $\ell$ of a constructor $c$ of sort $i$, then has
an unknown summary. Thus $u$ contains a recursive field $Y_j$ with unknown
$\mathcal E_j$, preceded only by constants and nullable recursive fields.
Because the family is inhabited, sort $i$ occurs at some node of a
finite value of a root sort. Replace the subtree at that node by
$c(\mathbf v)$, where $v_\ell$ is an element if $u$ is a terminal suffix and a
finite value containing an element if $u$ is a frame suffix; the fields of
$u$ before $Y_j$ hold element-free values, the field $Y_j$ holds $t_n$, and the
remaining fields hold finite values. Fix the surrounding context and all
finite fillers independently of $n$. In the terminal case, place the cursor at
$v_\ell$. In the frame case, place it at the last element of $v_\ell$ in
direction $d$; its move returns from $v_\ell$ after finitely many steps and
then runs the frame suffix~$u$.

Lemma~\ref{lem:context-reachability} ensures that entry and finitely many
moves reach this cursor. No ancestor of its focus lies inside $t_n$, so its
trail depth $h$ is fixed. Its next move crosses the element-free fields and
enters $t_n$, taking at least $n$ steps. Each pumped edge enters a new
constructor, so demands are unbounded too. Thus no function of $h$ bounds
these moves, proving the reverse implication of part~2.

\emph{Part 3.} Replace $t_n$ by $t_\infty$. The prefix reaching the cursor
visits only its finite ancestors and the fields preceding the focus in
direction $d$; it never demands the replaced subtree. The same entry and
moves reach the cursor on a productive infinite root; its next move diverges.
\end{proof}

\section{Path--trail correspondence proof}\label{app:proof-lem-path-factorisation}
\begin{proof}[Proof of Lemma~\ref{lem:path-factorisation}]
\emph{Setting.}
Write $P=\mu Q.\,(B+L\times Q\times R)$ for the paths, with constructors
$\mathsf{Stop}(b)$ and $\mathsf{Step}(l,p,r)$, where $b\in B$, $l\in L$,
$p\in P$, and $r\in R$. A history is a value of $\bwd{L\times R}$, built
from \code{BNil} and \code{Snoc}. Paths are finite because $P$ is a least
fixed point, and histories are finite by definition. The left-to-right
order of a path's fields is
\[
 \mathit{fields}(\mathsf{Stop}(b))=b,\qquad
 \mathit{fields}(\mathsf{Step}(l,p,r))=l,\mathit{fields}(p),r.
\]
For steps $(l_1,r_1),\ldots,(l_n,r_n)$, outermost first, and terminal $b$,
this is $l_1,\ldots,l_n,b,r_n,\ldots,r_1$. From the definitions of
Section~\ref{sec:arden},
\[
 \mathit{before}(\code{Snoc}\;H\;(l,r))=\mathit{before}(H),l,\qquad
 \mathit{after}(\code{Snoc}\;H\;(l,r))=r,\mathit{after}(H),
\]
and both are empty for \code{BNil}.

\emph{The map and its inverse.}
Define $\varphi:P\to\bwd{L\times R}\times B$ by
$\varphi(p)=\mathit{down}(\code{BNil},p)$, where
\[
\begin{aligned}
 \mathit{down}(H,\mathsf{Stop}(b))&=(H,b),\\
 \mathit{down}(H,\mathsf{Step}(l,p,r))&=\mathit{down}(\code{Snoc}\;H\;(l,r),\;p).
\end{aligned}
\]
Define $\psi:\bwd{L\times R}\times B\to P$ by
$\psi(H,b)=\mathit{up}(H,\mathsf{Stop}(b))$, where
\[
\begin{aligned}
 \mathit{up}(\code{BNil},p)&=p,\\
 \mathit{up}(\code{Snoc}\;H\;(l,r),\;p)&=\mathit{up}(H,\;\mathsf{Step}(l,p,r)).
\end{aligned}
\]
One recursive call of $\mathit{down}$ is a \emph{transfer}: it moves the
outermost step of the residual path, with both its fields, onto the
history. One recursive call of $\mathit{up}$ is the inverse transfer. The
function $\mathit{down}$ recurses on a smaller path and $\mathit{up}$ on a
smaller history, so both terminate.

\emph{Invariant.}
A state $(H,p)$ pairs a history with a residual path. Its \emph{reading} is
the sequence $\mathit{before}(H)$, $\mathit{fields}(p)$, $\mathit{after}(H)$.
The states $(H,\mathsf{Step}(l,p,r))$ and $(\code{Snoc}\;H\;(l,r),p)$ both
read as
\[
 \mathit{before}(H),\,l,\,\mathit{fields}(p),\,r,\,\mathit{after}(H).
\]
A transfer maps the first state to the second, and the inverse transfer
maps the second to the first. Hence both preserve the reading.

\emph{Order preservation.}
Let $\varphi(p)=(H,b)$. The run of $\mathit{down}$ starts in state
$(\code{BNil},p)$, whose reading is $\mathit{fields}(p)$, and ends in state
$(H,\mathsf{Stop}(b))$, whose reading is $\mathit{before}(H)$, $b$, $\mathit{after}(H)$.
By the invariant the two readings are equal, which is the order-preservation
property.

\emph{Mutual inverses.}
First, $\psi(\mathit{down}(H,p))=\mathit{up}(H,p)$ for all $H$ and $p$, by
induction on $p$. For $p=\mathsf{Stop}(b)$, both sides equal
$\mathit{up}(H,\mathsf{Stop}(b))$. For $p=\mathsf{Step}(l,p',r)$,
\[
 \psi(\mathit{down}(H,p))
 =\psi(\mathit{down}(\code{Snoc}\;H\;(l,r),p'))
 =\mathit{up}(\code{Snoc}\;H\;(l,r),p')
 =\mathit{up}(H,p),
\]
by the definitions of $\mathit{down}$ and $\mathit{up}$ and the induction hypothesis. Taking $H=\code{BNil}$
gives $\psi(\varphi(p))=\mathit{up}(\code{BNil},p)=p$.

Second, $\varphi(\mathit{up}(H,p))=\mathit{down}(H,p)$ for all $H$ and $p$,
by induction on $H$. For $H=\code{BNil}$, both sides equal
$\mathit{down}(\code{BNil},p)$. For $H=\code{Snoc}\;H'\;(l,r)$,
\[
 \varphi(\mathit{up}(H,p))
 =\varphi(\mathit{up}(H',\mathsf{Step}(l,p,r)))
 =\mathit{down}(H',\mathsf{Step}(l,p,r))
 =\mathit{down}(H,p).
\]
Taking $p=\mathsf{Stop}(b)$ gives
$\varphi(\psi(H,b))=\mathit{down}(H,\mathsf{Stop}(b))=(H,b)$. Hence
$\varphi$ is a bijection with inverse $\psi$.

\emph{Cost.}
Costs are structural work in the model of Section~\ref{sec:cost}. A
transfer matches one $\mathsf{Step}$ constructor, allocates one \code{Snoc}
cell and one pair, and copies the references $l$, $p$, and $r$. An inverse
transfer matches one \code{Snoc} cell and its pair, allocates one
$\mathsf{Step}$ cell, and copies the same three references. Each therefore
takes constant structural work. On a path of $n$ steps, $\varphi$ performs
$n$ transfers and one final $\mathsf{Stop}$ match, which allocates the
result pair. On a history of $n$ steps, $\psi$ allocates one
$\mathsf{Stop}$ cell, performs $n$ inverse transfers, and ends with one
\code{BNil} match. Each direction thus takes $O(n+1)$ structural work and
allocates $O(n+1)$ spine cells.

\emph{No stored value is inspected.}
Every clause of $\mathit{down}$ and $\mathit{up}$ matches only a path
constructor, a history constructor, or a pair. The definitions are
parametric in $L$, $R$, and $B$: each stored $l$, $r$, or $b$ is bound to a
variable and copied as a reference, so it is shared and never inspected.
The bounds cover structural work on the finite spine, excluding the cost
of exposing lazy source constructors.
\end{proof}

\section{Contract semantics and lowering}
\subsection{Ordinary cursor contracts}
\label{app:contract-semantics}
A contract for operation $o$ is a record
$\langle W_o,D_o,P_o,G_o\rangle$: a work expression, a demand descriptor,
the assumptions under which the contract holds, and an optional certificate
bounding work and constructor demands. Work expressions are
\[
 W ::= \gamma\mid \xi\mid W+W\mid \gamma\mathbin{\cdot}\xi\mid\max(W,W),
\]
where $\gamma$ is a natural constant and
$\xi\in\{h,q,N,M,\mathit{fuel}\}$. Ordinary cursor counts are exact;
fuelled drivers instead report the upper bound $\max(0,\mathit{fuel})$ for a
transition budget $\mathit{fuel}$. The fold backend reports $3N$ for a
completed eager run and distinguishes constructor demands from
client-algebra applications (Appendix~\ref{app:folds}). The JSON reports
represent work as an abstract syntax tree alongside a printed description, and
also record assumptions, demand descriptions, and certificates. Demand
descriptors are
\[
 D::=\epsilon\mid\mathsf{search}(d)\mid\mathsf{traverse}(d)
       \mid\mathsf{match}\mid\mathsf{algebra}.
\]
For cursor operations, $\epsilon$ means no source demand,
$\mathsf{search}(d)$ records constructor matches up to the next yield or
boundary in direction $d$, and $\mathsf{traverse}(d)$ records those of a
complete traversal. The fold descriptors $\mathsf{match}$ and
$\mathsf{algebra}$ distinguish source-constructor demands from client-algebra
applications; their semantics are given in Appendix~\ref{app:folds}.
The relation below concerns ordinary cursors; resumable-driver bounds are stated in Corollary~\ref{cor:scheduling}. Its result postconditions
are the positional guarantees of Section~\ref{sec:correctness}; $P_o$ records
the assumptions stated in Section~\ref{sec:cost}.

Label every reference and generated control step by either
$\mathsf{match}(c)$, when it exposes source constructor $c$, or $\epsilon$,
otherwise. Write
\[
  \langle o,s\rangle\overset{\,w,\delta\,}{\Longrightarrow}s'
\]
for a finite prefix of $w$ generated control steps of $o$ from state $s$, with source-label trace
$\delta$, and write
$\langle o,s\rangle\Downarrow^{w,\delta}z$ when such a prefix ends in public
result $z$. Cursor focus and replacement take zero \emph{control} steps:
their bounded record accesses are accounted for by the structural bound of Theorem~\ref{thm:cost}.

Work and demand descriptors denote properties of the reference machine of
Section~\ref{sec:correctness}, not of the generated run. The encoding of that
section keeps every tag and stored field, so it is injective and a generated state can be decoded. By Lemma~\ref{lem:mixed-operations}, every valid cursor encodes a yield. For an
operation $o$ and state $s$, let $\mathcal R_o(s)$ be the reference run from the
corresponding configuration: $\mathsf{Enter}_d(t,[\,])$ for entry at
root $t$, and $\mathsf{Advance}_d$ of the decoded yield for a move. It stops at
its first $\mathsf{Yield}$ or $\mathsf{End}$ and may be infinite. For
\code{traverse}, $\mathcal R_o(s)$ is the complete reference traversal to
$\mathsf{End}$; \code{focus}, \code{replace}, and \code{close} run no search,
and their $\mathcal R_o(s)$ is empty.

For a contract $\langle W,D,P,G\rangle$ of $o$, a natural-valued
valuation $\rho$ \emph{agrees} with $s$ when each of $h,N,M$ occurring in
$W$ or $G$ is assigned its corresponding finite depth or count. If $q$
occurs and $\mathcal R_o(s)$ is finite, $\rho(q)$ is its length; an infinite
run imposes no condition on $\rho(q)$. Unused variables impose no condition.
Thus a local move requires no finite size for the surrounding input, whereas
the complete-traversal contract requires finite $N$ and $M$.

Write $\llbracket W\rrbracket_\rho$ for the value of $W$ under $\rho$, using
ordinary arithmetic and maximum. The descriptor
$\llbracket\mathsf{search}(d)\rrbracket(s)$ is the label trace
of $\mathcal R_o(s)$; $\mathsf{traverse}(d)$ denotes the trace of the complete
reference traversal and $\epsilon$ the empty trace. The judgement
$\boldsymbol\Phi,o\models_\rho\langle W,D,P,G\rangle$ holds when every source state agreeing
with $\rho$ and satisfying $P$ has all three properties below. An operation satisfies a contract when the judgement holds for every $\rho$.
\begin{enumerate}
\item \emph{Partial correctness.} If
  $\langle o,s\rangle\Downarrow^{w,\delta}z$, then $\mathcal R_o(s)$ is finite,
  $w=\llbracket W\rrbracket_\rho$,
  $\delta=\llbracket D\rrbracket(s)$, and, for entry and moves, $z$ encodes
  the final configuration of $\mathcal R_o(s)$. Focus returns the selected
  element, replacement changes only that occurrence, and closing
  rebuilds the root.
\item \emph{Prefix correspondence.} For entry, moves, and complete traversal,
  the generated run and $\mathcal R_o(s)$ have prefixes of exactly the same
  finite lengths. Corresponding prefixes have the same labels and related
  configurations. Focus, replacement, and closing are not
  searches; their work and demand are specified by partial correctness above.
\item \emph{Conditional termination.} If $G$ is present and demanded roots are
  productive, $\mathcal R_o(s)$ is finite and its length and demand count satisfy
  the bounds of $G$.
\end{enumerate}
For an uncertified move, $W=q$ and $D=\mathsf{search}(d)$ thus assert that
the generated search has the length, trace, and result of the reference
search; they assert no bound on that length, and a diverging search has
exactly the finite prefixes of a diverging reference run.
The relation abstracts away from the time to expose a demanded source
constructor and from client-algebra work.

\begin{proof}[Proof of Theorems~\ref{thm:entry-contract}, \ref{thm:move-contract},
and~\ref{thm:cursor-contract}]
Lemma~\ref{lem:mixed-operations} and Theorem~\ref{thm:family-trail} supply the representation of every valid cursor.
For entry (Theorem~\ref{thm:entry-contract}) and moves
(Theorem~\ref{thm:move-contract}), Lemma~\ref{thm:simulation} relates one
labelled reference step to one emitted clause. Induction on prefix length proves prefix correspondence,
including divergent runs. A terminating prefix gives $W=q$, the reference
demand trace, and the corresponding result. Concatenating the entry prefix and the successive move prefixes
gives the complete-traversal correspondence; Theorem~\ref{thm:cost} gives its
$2N+M$ count. For Theorem~\ref{thm:cursor-contract}, focus and replacement take
zero control steps and inspect no source constructor. Closing takes $h+1$ counted steps without source demand, as in the proof of Theorem~\ref{thm:cost}. These three
operations therefore satisfy partial correctness with their empty reference
searches.
Theorem~\ref{thm:family-prefix} supplies conditional termination and the
reported bounds. These cases cover every row of the emitted report.
\end{proof}

\subsection{Container lowering}
\label{app:lowering-proof}
\label{sec:lowering}
The frontend of Section~\ref{sec:source} lowers each accepted list,
\code{Maybe}, \code{Either}, or declared unary container application to a
memoised auxiliary sort. The first occurrence of \code{[}$\tau$\code{]},
\code{Maybe} $\tau$, \code{Either} $\tau_1$ $\tau_2$, or $U\,\tau$ receives
a fresh sort whose ordered regular body is recursively lowered; equal
applications reuse it. Within a declared unary container, the internal name
\code{Self} denotes its recursive occurrence at the unchanged parameter and
becomes the fresh sort. The set of applications reached this way is the lowering closure; Section~\ref{sec:source} requires it to be finite.

The source declaration has its own signature $\Sigma$. Its sorts are the
declared types and the container applications in the finite closure, such as
\code{[Tree a]}; its constructors are the declared ones together with
\code{[]} and \code{(:)}, \code{Nothing} and \code{Just}, \code{Left} and
\code{Right}, and the constructors of each $U$, at those sorts. A field of
type \code{a} is an $X$ field, a family or container application is a $Y$
field of that sort, and any other type is a constant. Source values are
finite or infinite constructor trees over $\Sigma$: prefix-closed maps from
finite paths to constructors that respect sorts and arities, with element
and constant payloads at nonrecursive fields. This includes
infinite Haskell lists, but does not model arbitrary Haskell computations.
Let $\mathsf{lower}$ rename each sort and constructor of $\Sigma$ to its
lowered counterpart at every node, and let $\pi$ rename lowered demand
labels (Appendix~\ref{app:contract-semantics}) back to source constructors.

\begin{theorem}[Regular-container lowering]
\label{thm:lowering}
For every accepted regular declaration, $\mathsf{lower}$ is a
bijection from source values over $\Sigma$, finite or infinite, to
lowered-family values at the corresponding root sort. It preserves occurrence
paths, and hence both occurrence orders. The renaming extends to derivative
records and finite trails. For the reference machine of
Section~\ref{sec:correctness}, runs from corresponding configurations
correspond step by step, with demand labels related by $\pi$.
Finite lowering closure gives finite generated code.
\end{theorem}
\begin{proof}
Memoisation gives one lowered sort per sort of $\Sigma$ and one
specialised constructor per source constructor, with the same fields in the
same order and of the same kinds; \code{Self} becomes the sort of its own
container application. The renaming is therefore a bijection of signatures
that preserves arity, field order, field kind, and field sort. Applied at every
node of a path map it yields a lowered value, and the inverse renaming undoes
it; both act node by node, so infinite values need no limit argument. Paths keep
their field positions, so occurrences and their orders are preserved. A
derivative record retains the same tagged position and other fields under
renaming; applying this correspondence to each frame preserves finite trails
and their endpoint sorts. The reference machine inspects only the sort, tag,
field order, and field kinds of each node,
so its runs on $t$ and $\mathsf{lower}(t)$ agree up to renaming, and $\pi$
maps each lowered demand label to its source constructor. A finite closure gives finitely many sorts and constructors, hence finite generated code.
\end{proof}

\section{Proof sketches for cursor correctness}
\label{app:proof-thm-simulation}
\label{app:proof-cor-correctness}
\label{app:proof-lem-mixed-operations}
This appendix states the lemmas behind Theorem~\ref{thm:correctness} and
sketches their proofs. For finite values of single self-recursive datatypes,
the artifact's Lean development mechanises Lemma~\ref{thm:simulation} in both
directions; the theorems are \code{sim\_step}, \code{sim\_step\_rev},
\code{bsim\_step}, and, for matching the same source constructor,
\code{simulation\_labels} and \code{bsimulation\_labels} in
\code{mechanisation/}. The other lemmas, and simulation
for families, are proved here on paper.

\begin{proof}[Proof of Theorem~\ref{thm:correctness}]
Part 1 is Lemma~\ref{thm:simulation}. Part 2 is
Lemma~\ref{lem:mixed-operations}. Part 3 is
Lemmas~\ref{lem:navigation}, \ref{lem:reversal}, and~\ref{lem:replacement}.
\end{proof}

\begin{lemma}[Simulation]
\label{thm:simulation}
Fix an accepted family description and a running configuration
$\sigma\approx\theta$. If $\sigma\longrightarrow\sigma'$, then $\theta$
executes exactly one generated search clause, reaching $\theta'$ with
$\sigma'\approx\theta'$, and both steps demand the same source constructor, if
any. Every clause executed from $\theta$ arises in this way. Related
configurations have equal root and current sorts and store the same fields, so
encoded states are well typed.
\end{lemma}
\begin{proof}[Proof sketch of Lemma~\ref{thm:simulation}]
Fix $\sigma\approx\theta$ and consider the form of $\sigma$.

\emph{Entry}, $\sigma=\mathsf{Enter}_d(c(\mathbf v),\kappa)$. Then
$\theta=\seek_d(\lceil\kappa\rceil,c(\mathbf v))$ matches $c$, the only source
match of the step, and emits Equation~\eqref{eq:emit} at $e_d(c)$. Its three
cases are those of $\scan_d$. An $X$ field $\ell'$ gives
$\code{Just}(\mathsf{Cursor}(\lceil\kappa\rceil,b_{c,\ell'}(\widehat{\mathbf v}_{\ell'}),v_{\ell'}))$,
which is related to $\mathsf{Yield}(c,\mathbf v,\ell',\kappa)$. A $Y_{p'}$ field gives
\[
 \seek_d(\mathsf{Snoc}(\lceil\kappa\rceil,k_{c,\ell'}(\widehat{\mathbf v}_{\ell'})),v_{\ell'})
 =\seek_d(\lceil(c,\ell',\widehat{\mathbf v}_{\ell'})::\kappa\rceil,v_{\ell'}),
\]
related to the reference successor and with current sort $p'$ on both sides. No remaining
field gives $\rise_d(\lceil\kappa\rceil,c(\mathbf v))$.

\emph{Return}, $\sigma=\mathsf{Return}_d(t,(c,\ell,\widehat{\mathbf v}_\ell)::\kappa)$.
Then $\theta=\rise_d(\mathsf{Snoc}(\lceil\kappa\rceil,k_{c,\ell}(\widehat{\mathbf v}_\ell)),t)$.
The frame tag identifies $c$ and $\ell$, the record supplies
$\widehat{\mathbf v}_\ell$, and $t$ supplies $v_\ell$. The clause removes the frame,
restoring the parent sort, and emits the scan after $\ell$ on
$\widehat{\mathbf v}_\ell[\ell:=t]$, which is the reference successor; no source constructor is
matched. On $[\,]$, $\rise_d$ meets \code{BNil} and returns
\code{Nothing}, related to $\mathsf{End}$.

\emph{Advance}, $\sigma=\mathsf{Advance}_d(c,\mathbf v,\ell,\kappa)$. Then $\theta$ is the move from $\mathsf{Cursor}(\lceil\kappa\rceil,b_{c,\ell}(\widehat{\mathbf v}_\ell),v_\ell)$. The body tag supplies $c$ and $\ell$ and the focus supplies $v_\ell$, so the move emits the same scan after $\ell$ as the reference step. It matches only generated records, so neither step demands a source constructor.

The three running configurations call distinct generated functions, and
each function selects a unique clause by its outer tag. Every executed
clause therefore comes from the corresponding reference step, proving the
converse. Each
emitted record stores the fields of its indexed derivative summand, and
pushing or popping a frame changes the current sort of the trail exactly as it
changes that of the stack (Theorem~\ref{thm:family-trail}), so related states
have equal sorts and are well typed.
\end{proof}

\begin{lemma}[Mixed-operation preservation]
\label{lem:mixed-operations}
Let a cursor be obtained by entry. After any finite sequence of successful
\code{next}, \code{previous}, and \code{replace} operations, its trail and
terminal record encode the stack and local position of a yield, and
\code{close} reconstructs the represented root with all replacements in their
selected positions.
\end{lemma}
\begin{proof}[Proof sketch of Lemma~\ref{lem:mixed-operations}]
Entry produces the encoding of a yield by Lemma~\ref{thm:simulation}. Simulation preserves the encoded stack and terminal record after each
successful move, in either direction. Replacement changes only the selected
element. Induction on the operation sequence therefore preserves the
representation. Closing reconstructs its root by Theorem~\ref{thm:family-trail}.
\end{proof}

\begin{lemma}[Navigation]
\label{lem:navigation}
On a finite, fully defined input, entry selects the first or last element
occurrence, or reports that there are none. Each move selects the adjacent
occurrence in its direction, and reports a boundary exactly when there is none.
\end{lemma}
\begin{proof}[Proof sketch of Lemma~\ref{lem:navigation}]
Induction on the remaining fields of a constructor shows that the
reference scan visits its direct elements and recursive traversals in
the specified order. A continuation returns to the field immediately
after the completed child. Structural induction on a finite value then
gives the full order and termination. Lemma~\ref{thm:simulation} gives the same
result for generated code.
\end{proof}

\begin{lemma}[Reversibility]
\label{lem:reversal}
On a finite, fully defined input, a successful move followed by the opposite
move restores an equal value and the marked position.
\end{lemma}
\begin{proof}[Proof sketch of Lemma~\ref{lem:reversal}]
The trail and terminal tag of a cursor
determine the path of its occurrence: each frame $k_{c,\ell}$ contributes the
choice $(c,\ell)$ and the terminal $b_{c,\ell}$ the last one. A backward search from
that configuration scans the fields before $\ell$ at the current constructor and,
on each return, the fields before the saved position at the next ancestor. By
the definition of the orders it yields the greatest occurrence that precedes
the current one in forward order, which is the occurrence a forward move has
just left. The forward case is symmetric. Pointer identity may differ, because
returning through a constructor may allocate it again.
\end{proof}

\begin{lemma}[Replacement]
\label{lem:replacement}
On a finite, fully defined input, \code{replace} changes only the selected
occurrence.
\end{lemma}
\begin{proof}[Proof sketch of Lemma~\ref{lem:replacement}]
Replacement changes the selected field without inspecting it; each
reconstruction clause inserts that field in the position named by its
tag.
\end{proof}

\end{document}